\documentclass[aps,prc,twocolumn,showpacs,superscriptaddress,footinbib]{revtex4-1}

\usepackage{graphicx}
\usepackage{dcolumn}
\usepackage{bm}
\usepackage{enumerate}
\usepackage{amsmath}

\usepackage[normalem]{ulem}  

\begin{document}


\title{Emerging $\gamma$-soft-like spectrum in $^{196}$Pt in the SU3-IBM (I)}

\author{Tao Wang}
\email{suiyueqiaoqiao@163.com}
\affiliation{College of Physics, Tonghua Normal University, Tonghua 134000, People's Republic of China}

\author{Bing-cheng He}
\email{bhe@anl.gov}
\affiliation{Physics Division, Argonne National Laboratory, Lemont, Illinois 60439, USA}

\author{Chun-xiao Zhou}
\email{zhouchunxiao567@163.com}
\affiliation{College of Mathematics and Physics Science, Hunan University of Arts and Science, Changde 415000, People's Republic of China}

\author{Dong-kang Li}
\email{ldk667788@163.com}
\affiliation{College of Physics, Tonghua Normal University, Tonghua 134000, People's Republic of China}

\author{Lorenzo Fortunato}
\email{fortunat@pd.infn.it}
\affiliation{Dipartimento di Fisica e Astronomia  ``G.Galilei''-Universit\`{a} di Padova, via Marzolo 8, I-35131 Padova, Italy }
\affiliation{INFN-Sez.di Padova, via Marzolo 8, I-35131 Padova, Italy}

\date{\today}

\begin{abstract}
\textbf{Abstract:} Recently, it has been argued that a spherical-like spectrum emerges in the SU3-IBM, opening up new approaches to understand the $\gamma$-softness in realistic nuclei. In a previous paper, $\gamma$-softness with degeneracy of the ground and quasi-$\gamma$ bands was observed. In this paper, another special point connected with the middle degenerate point is discussed, which is found to be related with the properties of $^{196}$Pt. This emergent $\gamma$-softness has also been shown to be important for understanding the prolate-oblate asymmetric shape phase transition. The low-lying spectra, $B(E2)$ values and quadrupole moments in $^{196}$Pt are discussed showing that the new model can account for several observed features. This is the first part of the discussions on the $\gamma$-soft-like spectrum of $^{196}$Pt.

\end{abstract}

\maketitle

\section{Introduction}

Recently, an extension of the interacting boson model with \textrm{SU(3)} higher-order interactions (SU3-IBM for short) was proposed to describe the spherical-like $\gamma$-soft spectra in $^{110}$Cd \cite{Wang22}, to explain the puzzling $B(E2)$ anomaly \cite{Wang20,Zhang22},  to discuss the prolate-oblate asymmetric shape phase transition in Hf-Hg region including the $^{196}$Pt \cite{wang23}, and to provide an \textrm{E(5)}-like description for $^{82}$Kr \cite{zhou23}. \textrm{O(6)} higher-order interactions were found to be unable to explain the $B(E2)$ anomaly \cite{wangtao}.  These works imply that the \textrm{SU(3)} symmetry dominates the quadrupole deformation of a nucleus, and the $\gamma$-softness in realistic nuclei may be an emergent phenomenon, which has a deep relationship with the \textrm{SU(3)} symmetry.

\textbf{The Cd puzzle \cite{Heyde11} is the main motivation for proposing the SU3-IBM \cite{Wang22}.} \textbf{The normal states of} the Cd isotopes show a new \textbf{spherical-like $\gamma$-soft} rotational behavior \textbf{(the spectra are like the phonon excitation of the spherical shape while the transitional rates are like the $\gamma$-soft rotation)}, which is unexpected in standard nuclear structure studies \cite{Garrett10,Garrett12,Batchelder12,Heyde16,Garrett18,Garrett19,Garrett20}. The key observation is that there is in fact no the $0_{3}^{+}$ state at the three phonon level \cite{Batchelder12,Garrett19,Garrett20}. \textbf{This puzzle also occurs on other spherical nuclei, such as Te and Pd \cite{Garrett18}, so it is called spherical nucleus puzzle here.}\textbf{ In this puzzle, the phonon excitation of the spherical shape is questioned and refuted. Just as Heyde and Wood said: ``deformation has become a major observed excitation mode, making rigid spherical nuclei rather the exception'', ``a shift in perspective is needed: sphericity is a special case of deformation'' \cite{Heyde16}.} \textbf{In the SU3-IBM, with the help of SU(3) higher-order interactions, the similar spherical-like $\gamma$-soft spectra was really found, and the spherical shape was replaced by a special quadrupole deformation \cite{Wang22}.}

\textbf{Recently, another unexpected phenomenon was also observed, which is called B(E2) anomaly \cite{168Os,166W,172Pt,170Os}. In neutron-deficient nuclei $^{168}$Os, $^{166}$W, $^{172}$Pt, $^{170}$Os, a very small $B_{4/2}=B(E2; 4_{1}^{+}\rightarrow 2_{1}^{+})/B(E2; 2_{1}^{+}\rightarrow 0_{1}^{+})$ was found, which is much smaller than 1.0 (a single particle feature). In these nuclei, the energy ratio $E_{4/2}=E_{4_{1}^{+}}/E_{2_{1}^{+}}$ between the $4_{1}^{+}$ state and the $2_{1}^{+}$ state is larger than 2.0, so they show a collective excitation. In general, small $B_{4/2}$ value should not appear in the collective behaviors.} The B(E2) anomaly also rejects conventional theoretical explanations, including the interacting boson model (IBM-2) calculations based on the SkM$^{*}$ energy-density functional and the symmetry-conserving configuration mixing (SCCM) calculations \cite{168Os,166W,172Pt,170Os}.\textbf{ In the SU3-IBM, this anomaly can be also explained \cite{Wang20,Zhang22}. It should be noticed that this is still the only explanation as yet, and the B(E2) anomaly is also related to $\gamma$-softness \cite{Wang20}.}

\begin{figure}[tbh]
\includegraphics[scale=0.33]{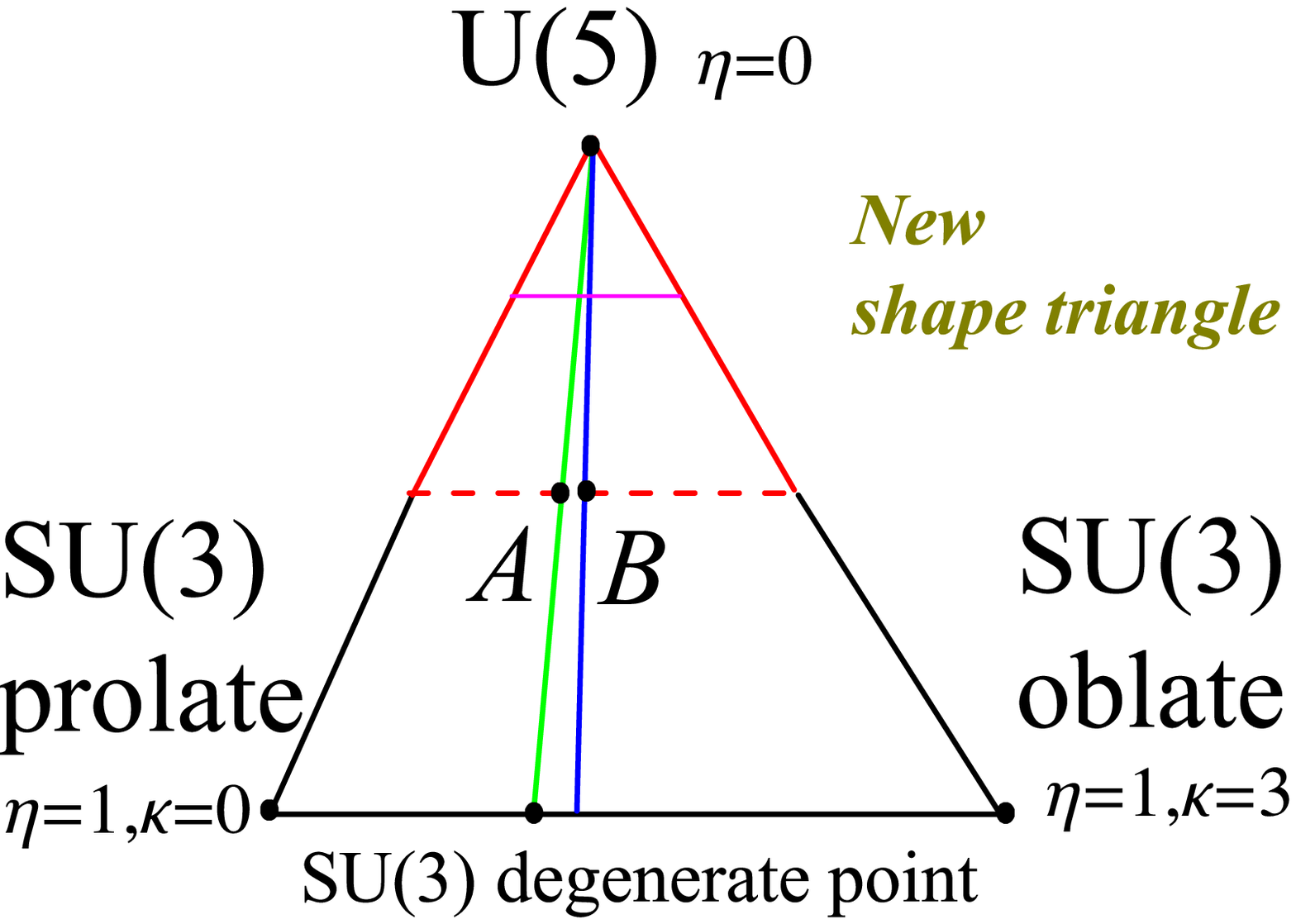}
\caption{New shape triangle: the top point of the triangle presents the \textrm{U(5)} limit, which is spherical shape. The two bottom points and the black line between them are all within the \textrm{SU(3)} limit.The left bottom point presents the \textrm{SU(3)} prolate shape, and the right one presents the \textrm{SU(3)} oblate shape. The middle point of the green line is point $A$, which is discussed in \cite{Wang22}. The middle point of the blue line is point $B$, which is discussed in this paper.}
\end{figure}

The successful explanation of these two abnormal phenomena makes the new theory of SU3-IBM very attractive, and further exploration of the applications of the theory becomes very valuable, especially on various $\gamma$-soft-like phenomena in nuclear spectra. What needs to be emphasized is that these two anomalous phenomena cannot be explained by the interacting boson model with the \textrm{O(6)} $\gamma$-softness \cite{wangtao}. Recent studies on the prolate-oblate asymmetric shape phase transition revealed that the key ingredient of the new model SU3-IBM is to describe the oblate shape with the SU(3) third-order interaction \cite{wang23}.
$\gamma$-softness comes from the competition between the prolate shape and the oblate shape, thus in the SU3-IBM the new $\gamma$-softness is an emergent phenomenon, which is different from the usual O(6) description \cite{Iachello78}.

\begin{figure}[tbh]
\includegraphics[scale=0.33]{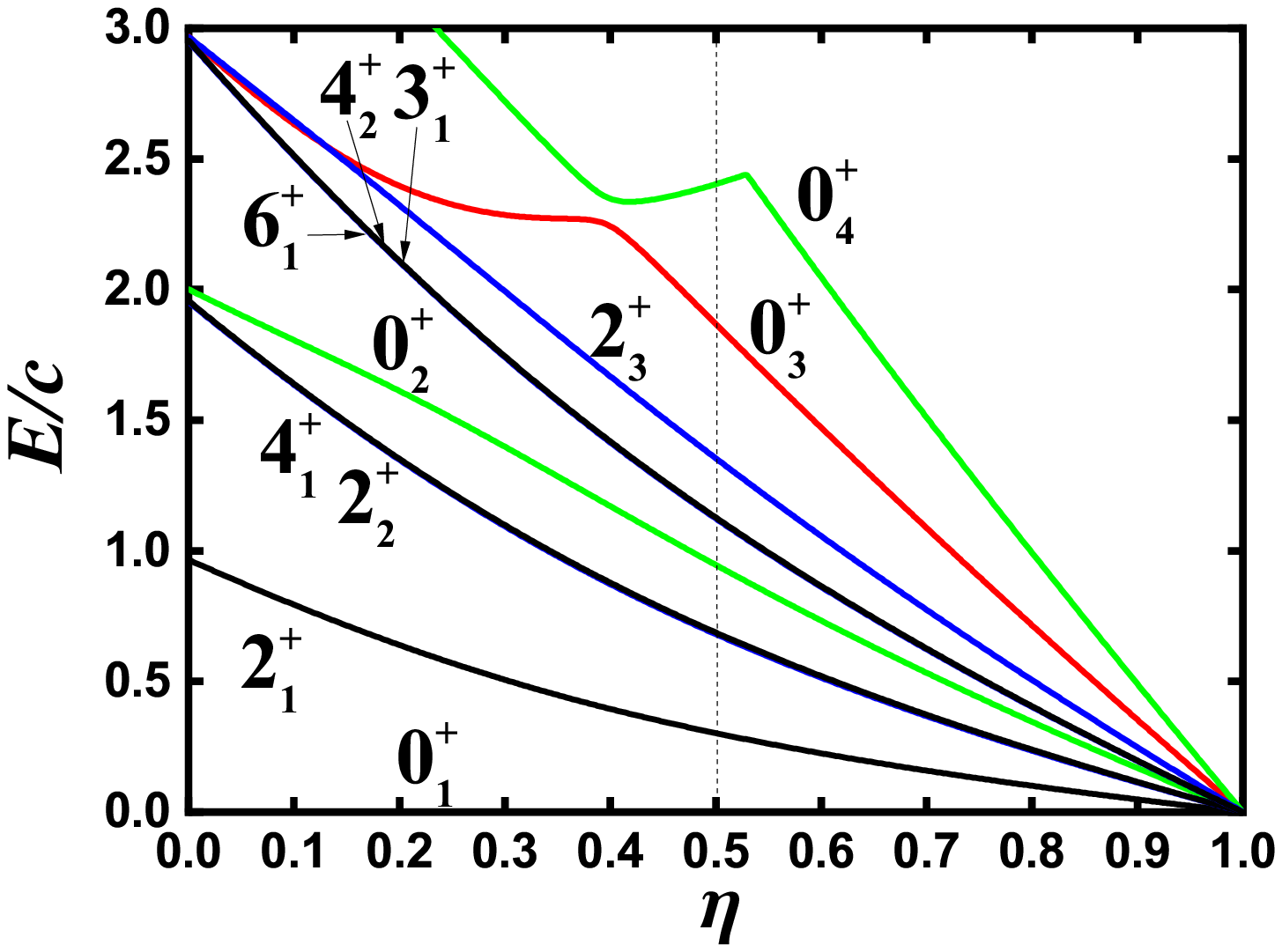}
\includegraphics[scale=0.33]{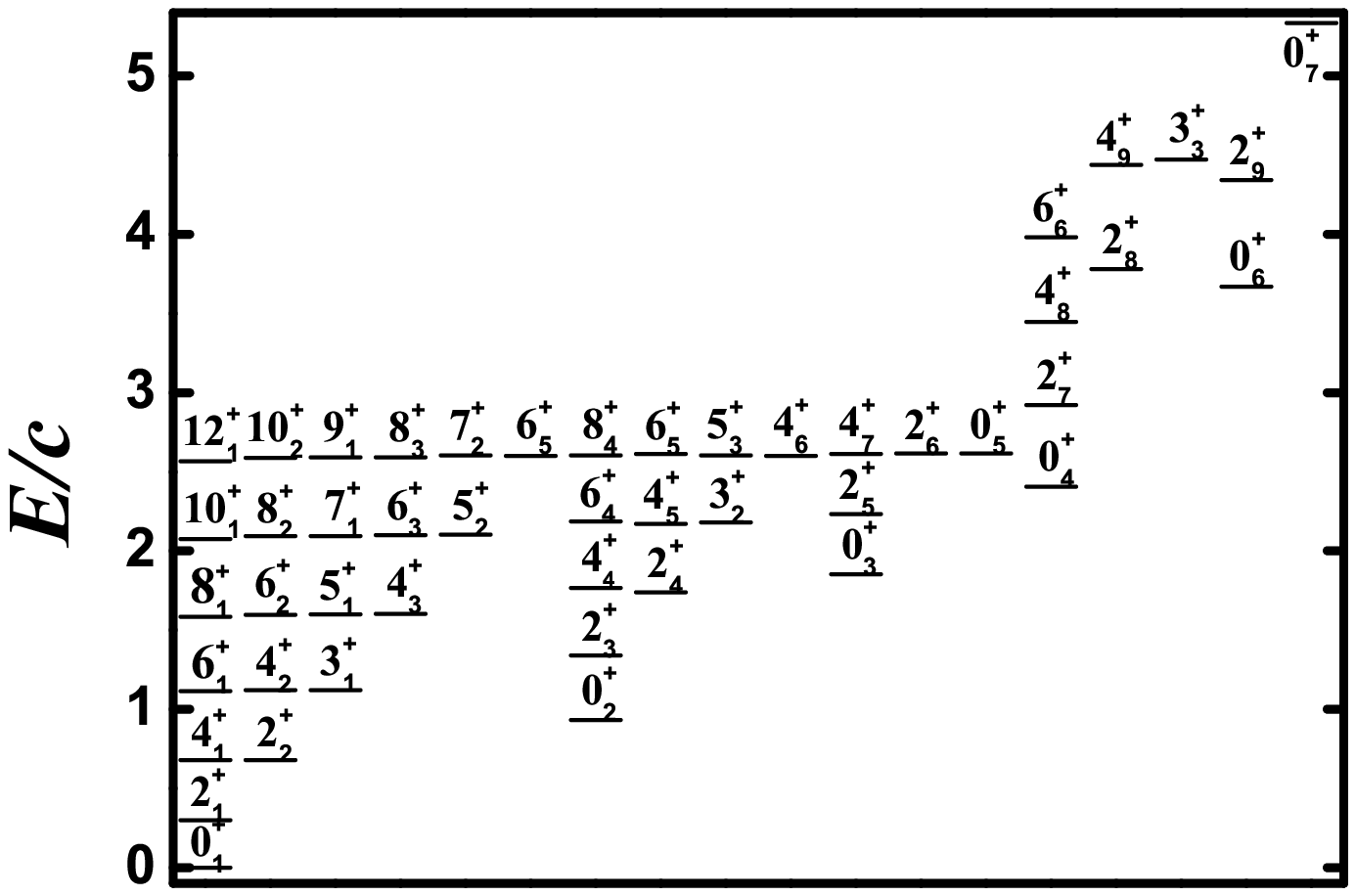}
\caption{Top: Partial low-lying level evolution along the green line in Fig. 1 for $N=6$. Bottom: Spherical-like spectra of the point $A$ for $N=6$.}
\end{figure}

The IBM provides an elegant approach to describe the low-lying collective excited behaviors in nuclear structure \cite{Iachello78,Iachello87}. In the simplest IBM-1, the basic building  constituents are the $s$ and $d$ bosons with angular momentum $l=0$ and $l=2$ respectively, and the collective states of a nucleus can be spanned by the su(6) algebra. Up to two-body interactions, a consistent-$Q$ (CQ) Hamiltonian  adopted in this model is \cite{Casten10,Wang08}
\begin{equation}
\hat{H}_{1}=c\left[(1-\eta) \hat{n}_{d}-\frac{\eta}{N}\hat{Q}_{\chi}\cdot \hat{Q}_{\chi}\right].
\end{equation}
 Here $\hat{n}_{d}=d^{\dag}\cdot \tilde{d}$ is the $d$-boson number operator,  $\hat{Q}_{ \chi}=[d^{\dag}\times\tilde{s}+s^{\dag}\times \tilde{d}]^{(2)}+\chi[d^{\dag}\times \tilde{d}]^{(2)} $ is the generalized quadrupole operator, $N$ is the total boson number, $c$ is a scale parameter and $0\leq \eta \leq 1, -\frac{\sqrt{7}}{2}\leq \chi \leq \frac{\sqrt{7}}{2}$ are parameters that allow to span a full range of different nuclear spectra. Although the formalism is simple, it can describe the spherical ($\eta=0$, the \textrm{U(5)} limit), prolate ($\eta=1$, $\chi=-\frac{\sqrt{7}}{2}$, the \textrm{SU(3)} limit), oblate ($\eta=1$, $\chi=\frac{\sqrt{7}}{2}$, the $\overline{\textrm{SU(3)}}$ case) and  $\gamma$-unstable ($\eta=1$, $\chi=0$, the \textrm{O(6)} limit) nuclei. This Hamiltonian is extensively used in fitting realistic nuclear spectra and discussing the shape phase transitions between different shapes \cite{Iachello87,Casten10,Wang08,Werner01,Jolie03}.

More than a decade ago, one of the authors (L. Fortunato) and his collaborators generalized the simple formalism (1), and a cubic-$Q$ interaction is introduced as follows \cite{Fortunato11}
\begin{small}
\begin{equation}
\hat{H}_{2}=c\left[(1-\eta) \hat{n}_{d}-\frac{\eta}{N}(\hat{Q}_{\chi}\cdot \hat{Q}_{\chi}+ \frac{\kappa_{3}}{N}[\hat{Q}_{\chi}\times \hat{Q}_{\chi}\times \hat{Q}_{\chi}]^{(0)})\right],
\end{equation}
\end{small}
where $\kappa_{3}$ is the coefficient of the cubic term. In the \textrm{SU(3)} limit, when $\chi=-\frac{\sqrt{7}}{2}$, the cubic interaction can describe an oblate shape (\textrm{SU(3)} oblate), which is different from the previous $\overline{\textrm{SU(3)}}$ oblate shape in Hamiltonian (1). This indicates that the previous description of the oblate shape with $\overline{\textrm{SU(3)}}$ symmetry can be replaced by the \textrm{SU(3)} symmetry and a new evolutional path from the prolate shape to the oblate shape can be established within only the \textrm{SU(3)} limit, see the bottom black line in Fig. 1. Thus an analytically solvable prolate-oblate shape phase transitional description within the \textrm{SU(3)} limit can be provided, see Ref. \cite{Zhang12}, which offers a rare example for finite-$N$ first-order quantum shape transition. The phase transitional point is also a degenerate point \cite{Zhang12}, which implies a hidden symmetry \cite{Wang22}. \textbf{This degenerate point is called SU(3) degenerate point, see the black point in the bottom black line in Fig. 1.} This hidden symmetry is responsible for the whole new progress in \cite{Wang22,Wang20,Zhang22,wang23,zhou23,wangtao}. Moreover, in this extended Hamiltonian $\hat{H}_{2}$, there is only a very tiny region of rigid triaxiality in the large-$N$ limit at $\chi=-\frac{\sqrt{7}}{2}$ when the parameter changes from the \textrm{U(5)} limit to the \textrm{SU(3)}  degenerate point, see the green line in Fig. 1.

These new results presented by Ref. \cite{Fortunato11,Zhang12} encourage us to understand the existing experimental phenomena from a new perspective. Some new and unexpected results have emerged recently. A new shape triangle can be drawn (see Fig. 1), which is similar to the Casten triangle related to the Hamiltonian (1) \cite{Casten06}. In the SU3-IBM new theory \cite{Wang22}, new $\gamma$-soft-like triaxial rotation is found, which is different from the \textrm{O(6)} $\gamma$-unrelated rotational mode in Hamiltonian (1). The shape transitional behaviors from the \textrm{U(5)} limit to the \textrm{SU(3)} degenerate point was numerically explored (green line in Fig. 1). Within the parameter region of the green line in Fig. 1, we find the unexpected result that there is an accidental degeneracy of the corresponding energy levels between the ground and quasi-$\gamma$ bands such that they form an exactly degenerate multiplet, \textbf{see the top graph in Fig. 2.} It will be clear, from the ensuing discussion, that while this degenerate multiplet corresponds to that found in the \textrm{SO(5)} symmetry with quantum number $\tau=2$, the next one ($\tau=3$) is not exactly degenerate, a feature that is often observed in actual nuclear spectra. The key observation is that, spherical-like $\gamma$-soft triaxial rotational spectra actually exists (\textbf{see the bottom figure in Fig. 2}), which may be the candidate to solve the spherical nucleus puzzle \cite{Heyde11,Heyde16,Garrett18}. \textbf{This spherical-like spectra is not shown in \cite{Wang22}.}

Historically, higher-order interactions in IBM-1 were introduced to  describe $\gamma$-rigid triaxial deformation and interactions $[d^{\dag}d^{\dag}d^{\dag}]^{(L)}\cdot[\tilde{d}\tilde{d}\tilde{d}]^{(L)}$ can play a key role for triaxiality of the ground state \cite{Isacker81,Isacker84}. An important progress related with our works is investigating \textrm{SU(3)} symmetry-conserving higher-order interactions \cite{Isacker85}. Subsequently, within the \textrm{SU(3)} limit, an algebraic realization of the rigid asymmetric rotor was established \cite{Isacker00, zhang14}. Recently, this realization has been used to explain the $B(E2)$ anomaly \cite{Zhang22}. \textrm{SU(3)} third-order and fourth-order interactions are also discussed in Ref. \cite{Rowe77,Draayer85,Elliott86,Kota20}. Higher-order terms are also important in partial dynamical symmetry \cite{Leviatan11}. Higher-order interaction $(\hat{Q}_{0}\times\hat{Q}_{0}\times\hat{Q}_{0})^{(0)}$ can present a rotational spectrum \cite{Isacker99}, where $\hat{Q}_{0}$ is the quadrupole operator in the \textrm{O(6)} limit. This result was further studied by \cite{Rowe05, Dai12}. However the O(6) symmetry was questioned in \cite{wangtao}. In these series of new developments \cite{Wang22,Wang20,Zhang22,wang23,zhou23,wangtao, Fortunato11,Zhang12,Isacker00,zhang14}, \textrm{SU(3)} higher-order interactions begin to show an extremely important role, albeit at a phenomenological level. These higher-order interactions have already been shown to be relevant to some realistic anomalies in nuclear structure \cite{Wang22,Wang20,Zhang22}, so introducing these terms is of practical significance.

The $\gamma$-soft shape was first described in Ref. \cite{Jean56}, where the geometric Hamiltonian is not dependent on the $\gamma$ variable. In the IBM, the $\gamma$-soft case can be described by the \textrm{O(6)} limit \cite{Iachello78,Casten78} and the nucleus of $^{196}$Pt was the first candidate for the \textrm{O(6)} spectra. However there was still some debates about it \cite{Fewell85,Casten87}. In the IBM-2 \cite{Iachello87}, triaxial shape can be described even with up to two-body interactions \cite{Bijker82,Dukelsky04,Caprio04,Caprio05}. Three-body interactions $[d^{\dag}d^{\dag}d^{\dag}]^{(L)}\cdot[\tilde{d}\tilde{d}\tilde{d}]^{(L)}$ are also used in the IBM-2 to investigate the $\gamma$ triaxiality \cite{Nomura12}. In the sdg-IBM, $l=4$ g bosons can be introduced and hexadecapole deformation can be discussed \cite{Isacker10}. Except for the IBM, triaxial shapes are also investigated by many existing nuclear models \cite{Casten10,Heyde11,Bohr75,Ring80,Reinhard03,Zuker05,Ring11,Shimizu12,Nomura21}.

Although $^{196}$Pt seems to adapt to the description in terms of the \textrm{O(6)} symmetry, some noticeable deviations still exist and cannot be described at a satisfactory level in the IBM. The first drawback is that it has a large electric quadrupole moment \cite{Fewell85}, $Q_{2_{1}^{+}}$=0.62(8), pointing towards the oblate side. The second is that the staggering feature of  $\gamma$ band breaks the \textrm{O(5)} symmetry, which seems to be intermediate between the $\gamma$-soft and $\gamma$-rigid. The third is the positions of the $0_{2}^{+}$, $0_{3}^{+}$, $0_{4}^{+}$ states, which can not be reproduced well \cite{Leviatan09}.

Recently two important results in the SU3-IBM have also been found \cite{wang23,zhou23}. In \cite{wang23}, the SU(3)-IBM is used to describe the prolate-oblate shape phase transition with an asymmetric way, which can well explain the shape transitions from $^{180}$Hf to $^{200}$Hg including the nucleus $^{196}$Pt. \textbf{The transitional region is chosen along the dashed red line in Fig. 1, which passes through the middle point $A$ of the transitonal region from the U(5) limit to the SU(3) degenerate point.} It was found that, in this shape phase transition, another special point that shows accidental degeneracy features, can be located near the middle of the degenerate line \textbf{(point $B$ in Fig. 1)} and it can be used to describe the properties of $^{196}$Pt \cite{wang23}. In \cite{zhou23}, a shape transitional behavior like the one from the \textrm{U(5)} limit to the \textrm{O(6)} limit in IBM is found by introducing the \textrm{SU(3)} fourth-order interaction, which can describe the \textrm{E(5)}-like $\gamma$-softness in $^{82}$Kr.

Following the ideas of the previous researches, further exploration of the applications of the SU3-IBM is necessary. The $^{196}$Pt is the focus. The three obvious deficiencies discussed above can be well overcome simultaneously. This nucleus is usually regarded as a typical example with \textrm{O(6)} symmetry in Hamiltonian (1). We have found that the new model can give a more reasonable description and our work provides a new understanding for $\gamma$-softness in $^{196}$Pt and other similar nuclei, which is related with the SU(3) symmetry. Thus these results in the SU3-IBM (\cite{Wang22,Wang20,Zhang22,wang23,zhou23,wangtao} and this paper) together confirm the validity of the new idea.

\textbf{This is the first paper in which the nucleus $^{196}$Pt is discussed in terms of the SU(3) higher order interactions.}
In the SU(3) limit, three interactions ( the SU(3) second-order Casimir operator $\hat{C}_{2}[\textrm{SU(3)}]$, the SU(3) third-order Casimir operator $\hat{C}_{3}[\textrm{SU(3)}]$ and the square of the second-order Casimir operator $\hat{C}_{2}^{2}[\textrm{SU(3)}]$ ) can determine the quadrupole shape of the ground state and the energies of the $0^{+}$ states \cite{zhou23}. In this paper, the three operators are first investigated to fit the lowest few $0^{+}$ states in $^{196}$Pt. Other three operators $[\hat{L} \times \hat{Q} \times \hat{L}]^{(0)}$, $[(\hat{L} \times \hat{Q})^{(1)} \times (\hat{L} \times \hat{Q})^{(1)}]^{(0)}$ and $\hat{L}^{2}$ can describe the triaxial rotational spectra \cite{Isacker00,zhang14}, which will be discussed in the series of following papers.

\section{Hamiltonian}

In the SU3-IBM, the $d$ boson number operator $\hat{n}_{d}$ must be included, which can describe a spherical shape. This is vital for the pairing interaction between the valence nucleons. Other interacting terms are all \textrm{SU(3)} conserving invariants. The traditional second-order interaction $-\hat{C}_{2}[\textrm{SU(3)}]$ can describe the prolate shape. Ref. \cite{Fortunato11} pointed out that the third-order Casimir operator $\hat{C}_{3}[\textrm{SU(3)}]$ can describe the oblate shape. Other higher-order interactions should be considered in some peculiar phenomena, such as $B(E2)$ anomaly and some unusual experimental data which can not be described by previous theories \cite{Wang20,Zhang22}. In \cite{zhou23}, the square of the second-order Casimir operator $\hat{C}_{2}^{2}[\textrm{SU(3)}]$ is found to be vital for the $\gamma$-softness of the realistic nuclei. In this paper, the third-order invariant operator and the square of the second-order invariant operator are introduced into the interactions, like in Ref. \cite{zhou23} (the fourth-order interaction is only a supplementary term here). Although this is a simple formalism in the SU3-IBM, it shows many new interesting phenomena. Thus the Hamiltonian discussed in this paper is
\begin{eqnarray}
  \hat{H}&=&c\left[(1-\eta)\hat{n}_{d}+\eta\left(-\frac{\hat{C}_{2}[\textrm{SU(3)}]}{2N}\right.\right. \nonumber \\&&
  \left.\left.+\kappa \frac{\hat{C}_{3}[\textrm{SU(3)}]}{2N^{2}}+\xi\frac{\hat{C}_{2}^{2}[\textrm{SU(3)}]}{2N^{3}} \right)\right],
\end{eqnarray}
where $0\leq \eta \leq 1$, \emph{c} is a global energy scale parameter,\emph{ N} is the boson number, $\kappa$ is the coefficient of the cubic term, $\kappa=\frac{9\kappa_{_{3}}}{2\sqrt{35}}$, $\xi$ is the coefficient of the fourth-order interaction, and $\hat{C}_{2}[\textrm{SU(3)}]$ and $\hat{C}_{3}[\textrm{SU(3)}]$ are the second-order and third-order \textrm{SU(3)} Casimir operators separately. If the fourth-order term is not considered, \textbf{that is $\xi=0.0$}, the Hamiltonian (3) can be described by the new shape triangle in Fig. 1.

\begin{figure}[tbh]
\includegraphics[scale=0.33]{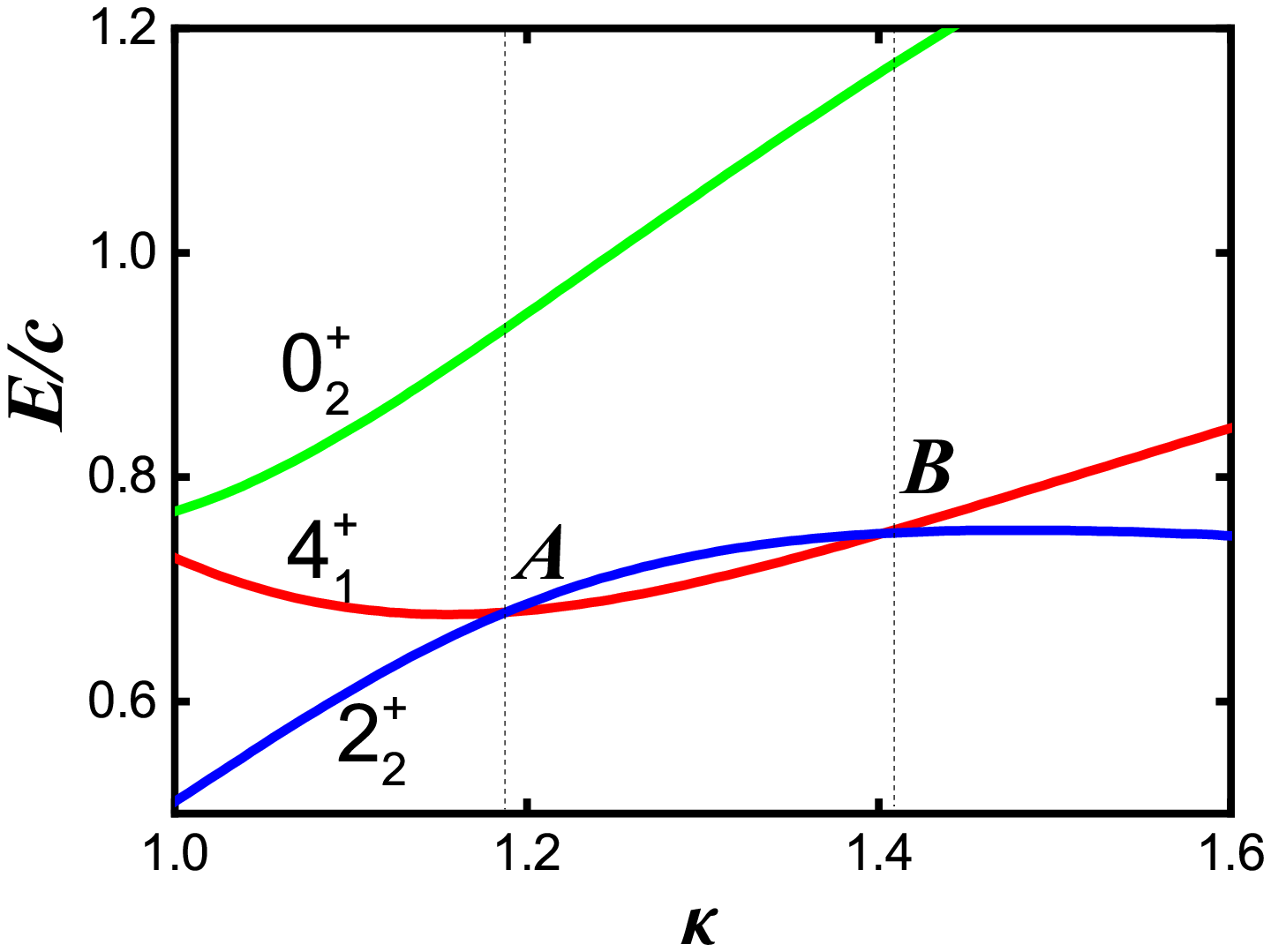}
\caption{Level evolution of the $4_{1}^{+}$, $2_{2}^{+}$ and $0_{2}^{+}$ states for the parameter $\kappa$ from 1.0 to 1.6 when $\eta=0.5$, $\xi=0$ and $N=6$. Two crossing points $A$ and $B$, where accidental degeneracy occurs, can be observed.}
\end{figure}

\textbf{Figure 1 requires some explanations. As introduced earlier, the top point of the new triangle in Fig. 1 is the U(5) limit ($\eta=0.0$), which comes from the pair interaction between nucleons and can introduce the spherical shape and the phonon excitation. Spherical nucleus puzzle implies that this limit can not exist alone. If $\eta=1.0$, it is the SU(3) limit and corresponds to the two bottom points and the black line between them. The left bottom point presents the SU(3) prolate shape ($\eta=1.0,\kappa=0.0$), and the right one presents the SU(3) oblate shape ($\eta=1.0,\kappa=3.0$). Thus the bottom black line presents the shape transition between the prolate shape to the oblate shape \cite{Zhang22}. These quadrupole shapes can be described by the SU(3) irrep $(\lambda,\mu)$. For a certain boson number $N$, the prolate shape is (2$N$,0) while the oblate one is (0,$N$). Thus the spectra of the prolate and oblate shapes are not the same, which is different from the old IBM with $\overline{\textrm{SU(3)}}$ symmetry.}

\textbf{The shape critical point between the prolate shape and the oblate shape is the SU(3) degenerate point \cite{Wang22,Zhang12}, which is denoted by the black point in the bottom black line. At the left side of this degenerate point, the shape is prolate while at the right side, it is oblate. This degenerate point inspires the SU3-IBM. The green line between the U(5) limit to the SU(3) degenerate point is the degenerate line where the $4_{1}^{+}$ state and the $2_{2}^{+}$ state are degenerate, see the top graph of Fig. 2. In fact, the green line is not straight, and it is a curved line. However the straight line is a good approximation if $N$ is small. The top part of the purple line (near $\eta=0.2$) is spherical, and the bottom part ($0.2 < \eta \leq 1.0$) presents the  deformations. The middle point $A$ of the green line is in the deformation region. We found that this point $A$ has the spherical-like spectra, see the bottom figure of Fig. 2. The dashed red line passes through the point $A$, which was discussed in \cite{wang23} to investigate the asymmetric shape phase transition from the prolate shape to the oblate shape in the Hf-Hg nuclei region. It was found that, along the dashed red line, there exists another accidental degenerate point $B$ between the $4_{1}^{+}$ state and the $2_{2}^{+}$ state, see Fig. 3. And interestingly, $^{196}$Pt was found to locate at this point $B$. The blue line begins from the U(5) limit and passes through the point $B$, which is the transitional region discussed in this paper.}

In the \textrm{SU(3)} limit the two Casimir operators can be related with the quadrupole second or third-order interactions as following \cite{Kota20,zhang14}
\begin{equation}
\hat{C}_{2}[\textrm{SU(3)}]=2\hat{Q}\cdot \hat{Q}+\frac{3}{4} \hat{L}\cdot \hat{L},
\end{equation}
\begin{small}
\begin{equation}
\hat{C}_{3}[\textrm{SU(3)}]=-\frac{4}{9}\sqrt{35}[\hat{Q}\times \hat{Q} \times \hat{Q}]^{(0)}-\frac{\sqrt{15}}{2}[\hat{L}\times \hat{Q} \times \hat{L}]^{(0)},
\end{equation}
\end{small}
here  $\hat{Q}=[d^{\dag}\times\tilde{s}+s^{\dag}\times \tilde{d}]^{(2)}-\frac{\sqrt{7}}{2}[d^{\dag}\times \tilde{d}]^{(2)}$ is the SU(3) quadrupole operator and $\hat{L}=\sqrt{10}[d^{\dag}\times \tilde{d}]^{(1)}$ is the angular momentum operator. For a given \textrm{SU(3)} irrep $(\lambda,\mu)$, the eigenvalues of the two Casimir operators under the group chain $\textrm{U(6)}\supset \textrm{SU(3)} \supset \textrm{O(3)}$ are given as
\begin{equation}
\langle \hat{C}_{2}[\textrm{SU(3)}]\rangle=\lambda^{2}+\mu^{2}+\lambda \mu+3\lambda+3\mu,
\end{equation}
\begin{equation}
\langle \hat{C}_{3}[\textrm{SU(3)}]\rangle=\frac{1}{9}(\lambda-\mu)(2\lambda+\mu+3 ) (\lambda+2\mu+3 ).
\end{equation}

If $\kappa=\frac{3N}{2N+3}$, the second term in Hamiltonian (3) describes the \textrm{SU(3)} degenerate point ($\xi=0$). It should be noticed that the location of the \textrm{SU(3)} degenerate point along the variable $\kappa$ is related to the boson number $N$ \cite{wang23}. For $^{196}$Pt, $N=6$, it is found at $\kappa =1.2$. In the large-$N$ limit, $\kappa \rightarrow 1.5$. At this degenerate point, the \textrm{SU(3)} irreps satisfying the condition $\lambda+2\mu=2N$ are all degenerate.

\textbf{The Hamiltonian (3) is diagonalized using our SU(3) basis diagonalization Fortran code \cite{Wang08} with the $U(6)\supset SU(3) \supset SO(3)$ basis spanned by ${ | N(\lambda,\mu)\chi L \rangle}$, where $\chi$ is the branching multiplicity occurring in the reduction of $SU(3)\downarrow SO(3)$. The basis vectors are orthonormal, so the eigenstates can be expressed as
\begin{small}
\begin{equation}
\label{eq.2}
|N,L_{\zeta};\eta,\kappa,\xi\rangle
=\sum_{(\lambda,\mu)\chi}C^{L_{\zeta}}_{(\lambda,\mu)\chi}(\eta,\kappa,\xi)| N(\lambda,\mu)\chi L \rangle,
\end{equation}
\end{small}
where $\zeta$ is an additional quantum number distinguishing different eigenstates with the same angular momentum $L$ and $C^{L_{\zeta}}_{(\lambda,\mu)\chi}(\eta,\kappa,\xi)$ is the corresponding expansion coefficient. We plan to publish this Fortran code in a forthcoming paper.}

\begin{figure}[tbh]
\includegraphics[scale=0.33]{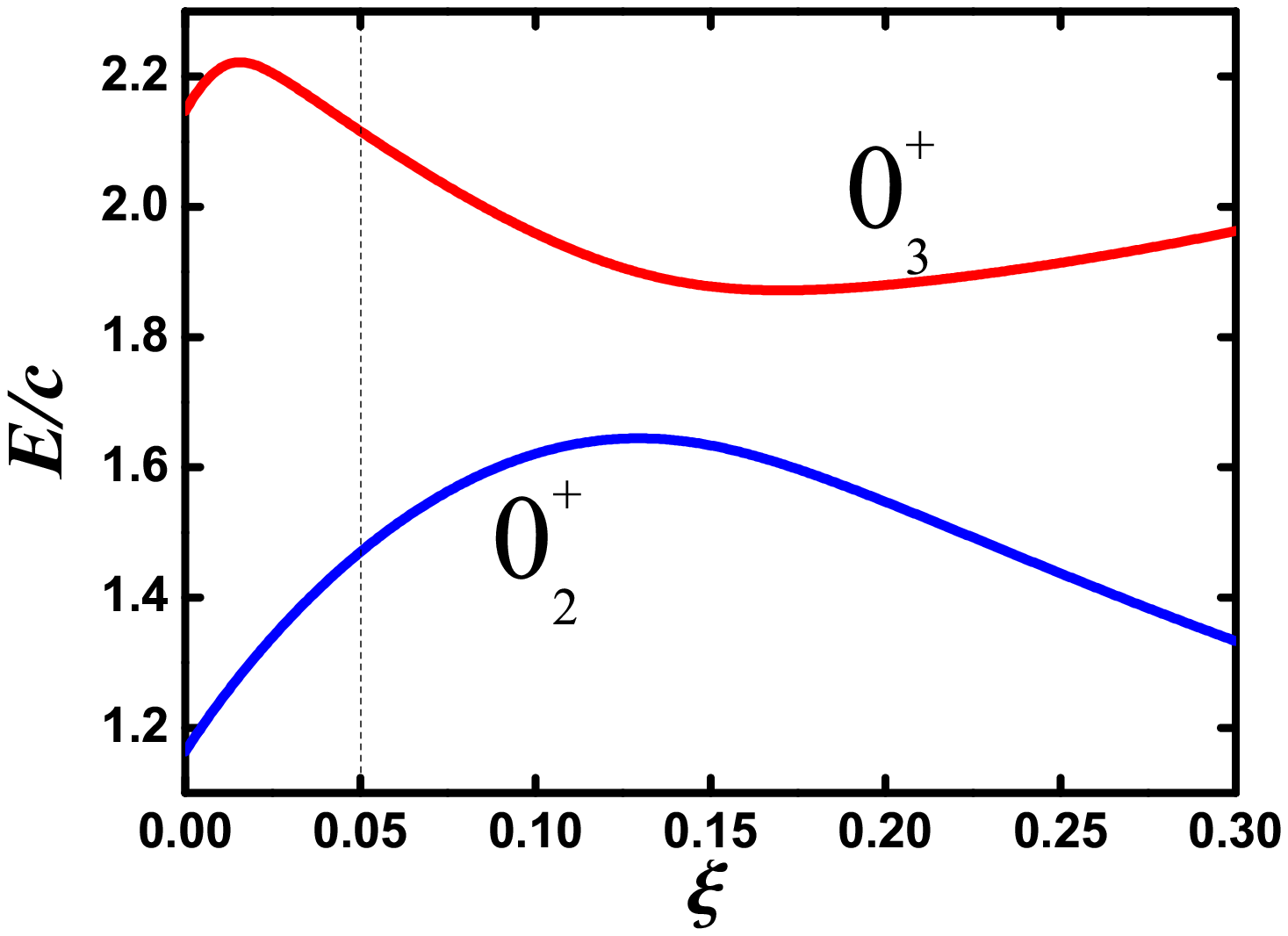}
\caption{Level evolution of the $0_{2}^{+}$ and $0_{3}^{+}$ states for the parameter $\xi$ from 0 to 0.30 when $\eta=0.5$, $\kappa=1.404$ and $N=6$.}
\end{figure}

\begin{figure}[tbh]
\includegraphics[scale=0.33]{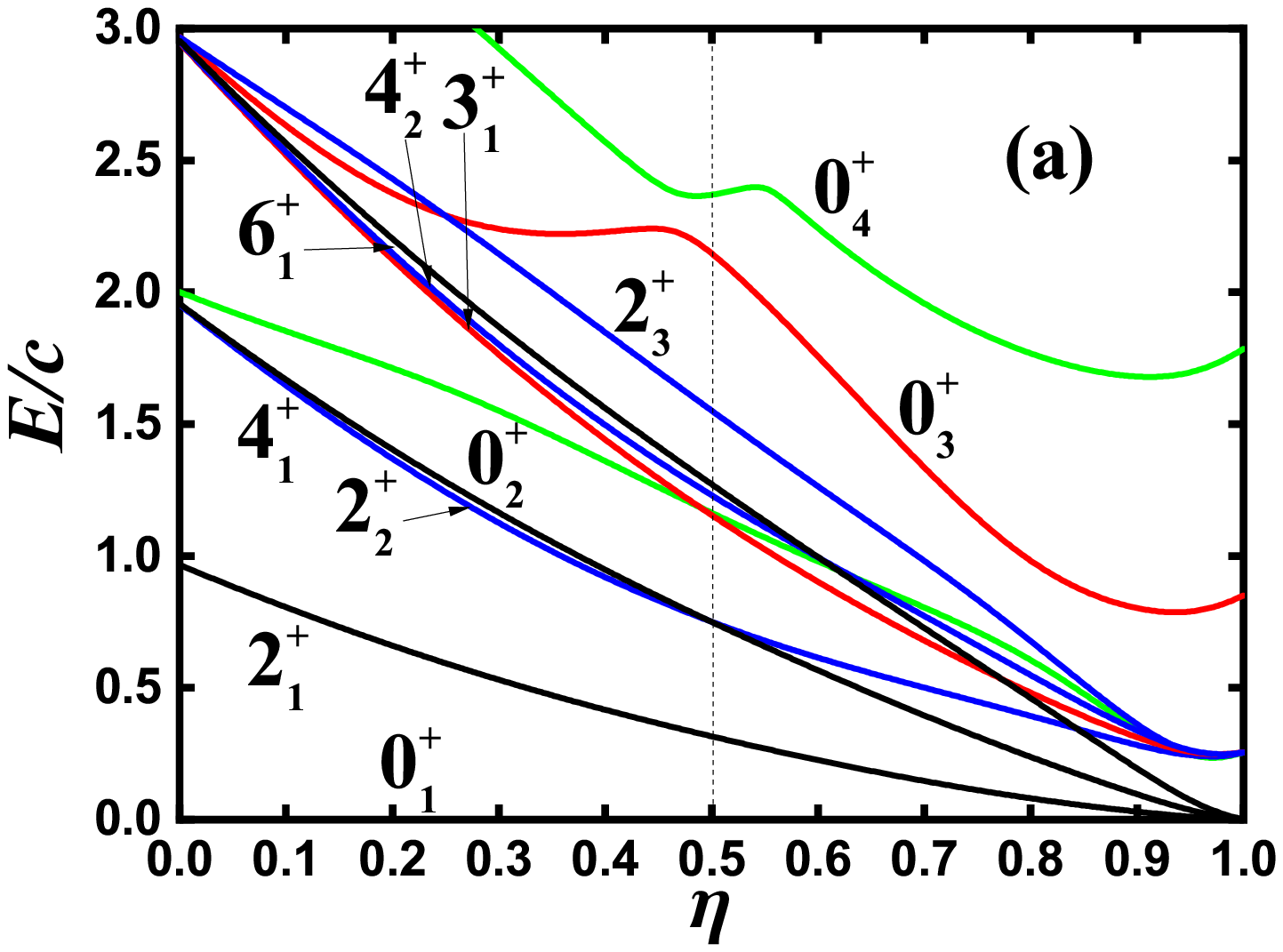}
\includegraphics[scale=0.33]{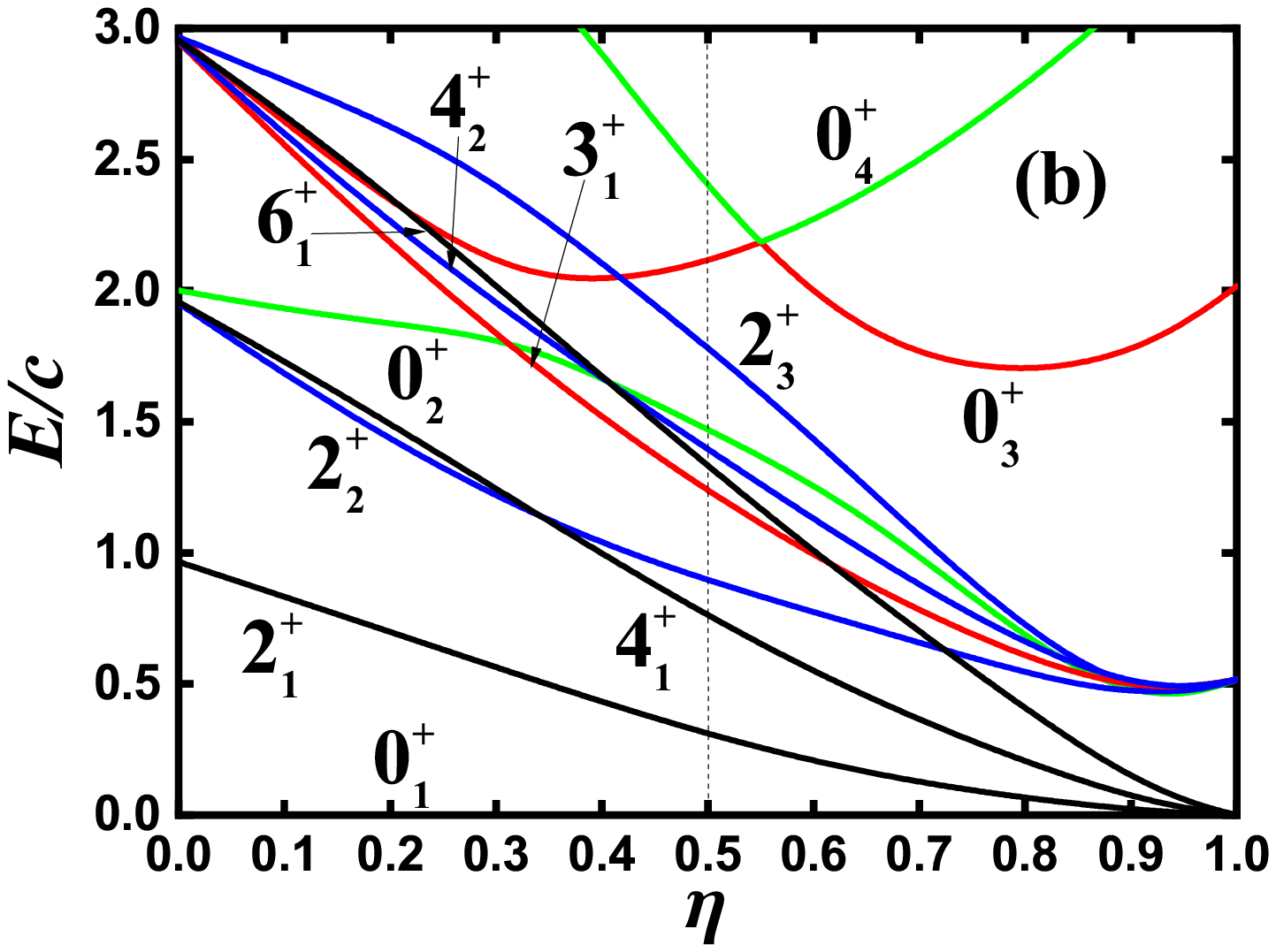}
\caption{Partial low-lying level evolution along the blue line in Fig. 1 for $N=6$ when (a) $\xi=0$ and (b) $\xi=0.05$.}
\end{figure}

\begin{figure}[tbh]
\includegraphics[scale=0.33]{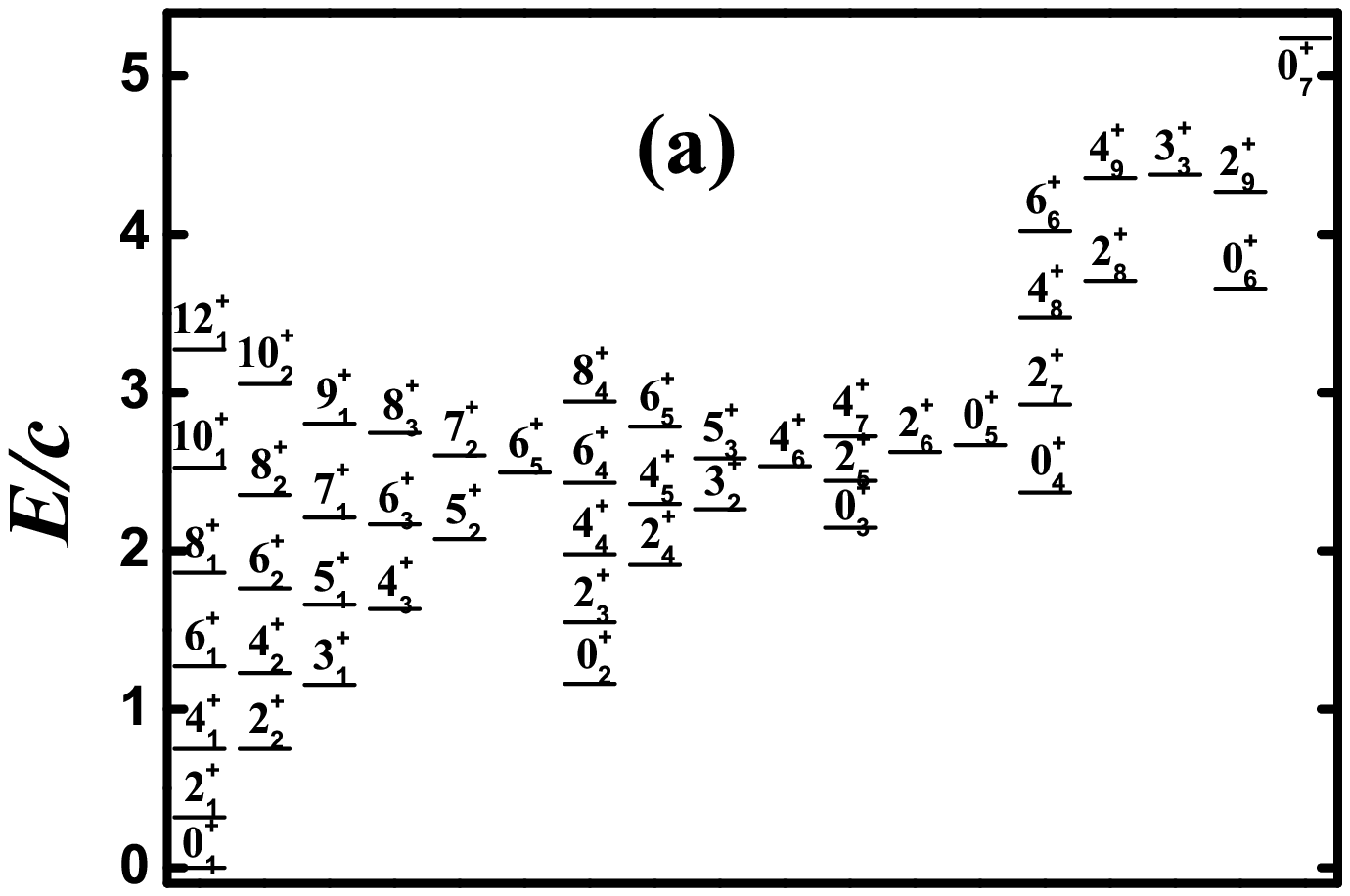}
\includegraphics[scale=0.33]{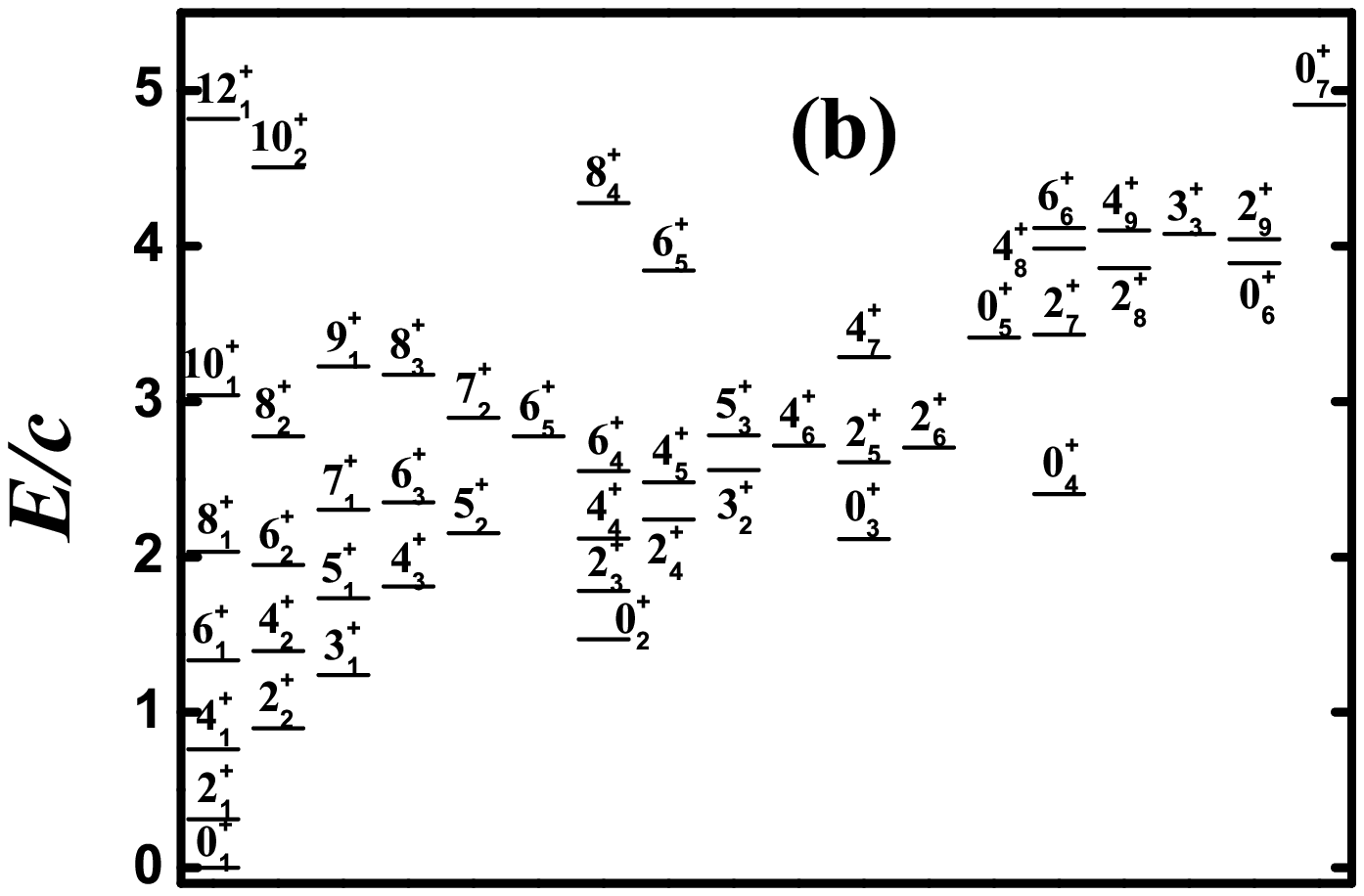}
\caption{Energy spectra of (a) the point $A$, (b) the cousin point $B$ and (c) adding the fourth-order interaction when $\xi=0.05$ for $N=6$.}
\end{figure}

\section{$\gamma$-soft-like spectra for the point $B$}

\textbf{The top figure of} Fig. 2 presents the partial low-lying level evolutions from the \textrm{U(5)} limit to the \textrm{SU(3)} degenerate point for $N=6$ as a function of $\eta$. The choice of the boson number $N=6$ correponds to $^{196}$Pt. (In the previous paper \cite{Wang22}, $N=7$ is discussed for the $^{110}$Cd.) It's clear that the four lowest $0^{+}$ states are all degenerate if $\eta=1.0$. The key finding is that the $4_{1}^{+}$ state and the $2_{2}^{+}$ state are degenerate, as well as the triplet  states $6_{1}^{+}$, $4_{2}^{+}$, $3_{1}^{+}$. This degeneracy can hold for some higher levels. Unfortunately the reason for this degeneracy is still unknown. This unexpected $\gamma$-softness is found in Ref. \cite{Wang22}.

The SU(3) degenerate point is found at $\kappa=1.2$. A further exploration around this degenerate point is undoubtedly useful for understanding the new $\gamma$-softness.  Fig. 3 plots level evolution of the $4_{1}^{+}$, $2_{2}^{+}$ and $0_{2}^{+}$ states for the parameter $\kappa$ from 1.0 to 1.6 when $\eta=0.5$, $\xi=0$  and $N=6$. Obviously there are two crossing points between the $4_{1}^{+}$ and $2_{2}^{+}$ states at $\kappa_{A}=1.188$ and $\kappa_{B}=1.404$. The position relationships of these three states are very important for understanding the $\gamma$-softness in realistic nuclei. The left one is the point $A$, and the right point is \textbf{point} $B$, which is the special point discussed in this paper.  The location of the point $B$ is $\kappa_{B}=1.404$. Point $A$ is biased toward the prolate side and point $B$ toward the oblate side. In the $\gamma$-soft-like region between the two points, point $B$ is closest to the oblate shape. $^{196}$Pt is a $\gamma$-soft nucleus with large positive quadruple moment, so it is natural to investigate whether the spectrum at point $B$ can be used to describe this nucleus. It should be noticed that the value of $\kappa_{A}=1.188$ is somewhat smaller than the value of the \textrm{SU(3)} degenerate point 1.2, thus the real degenerate line between the \textrm{U(5)} limit and the \textrm{SU(3)} degenerate point is not the directly connected green line in Fig. 1 \cite{Fortunato11}, which is also discussed in \cite{wang23}. There is a sudden shape change through the \textrm{SU(3)} degenerate point from the prolate shape to the oblate shape \cite{Zhang12}. However in the large-$N$ limit \cite{Fortunato11}, the point $A$ is the critical point between the prolate shape and the $\gamma$-rigid triaxial shape, while the point $B$ is the critical point between the $\gamma$-rigid triaxial shape and the oblate shape. Thus the positions of these points are different from each other \cite{wang23}. However, for small $N$, this connected line is a good approximation \cite{wang23}. In Fig. 1, the blue line passes through the point $B$. It should be noticed that, for any $N\geq 4$, this special point $B$ exists.

\begin{figure}[tbh]
\includegraphics[scale=0.33]{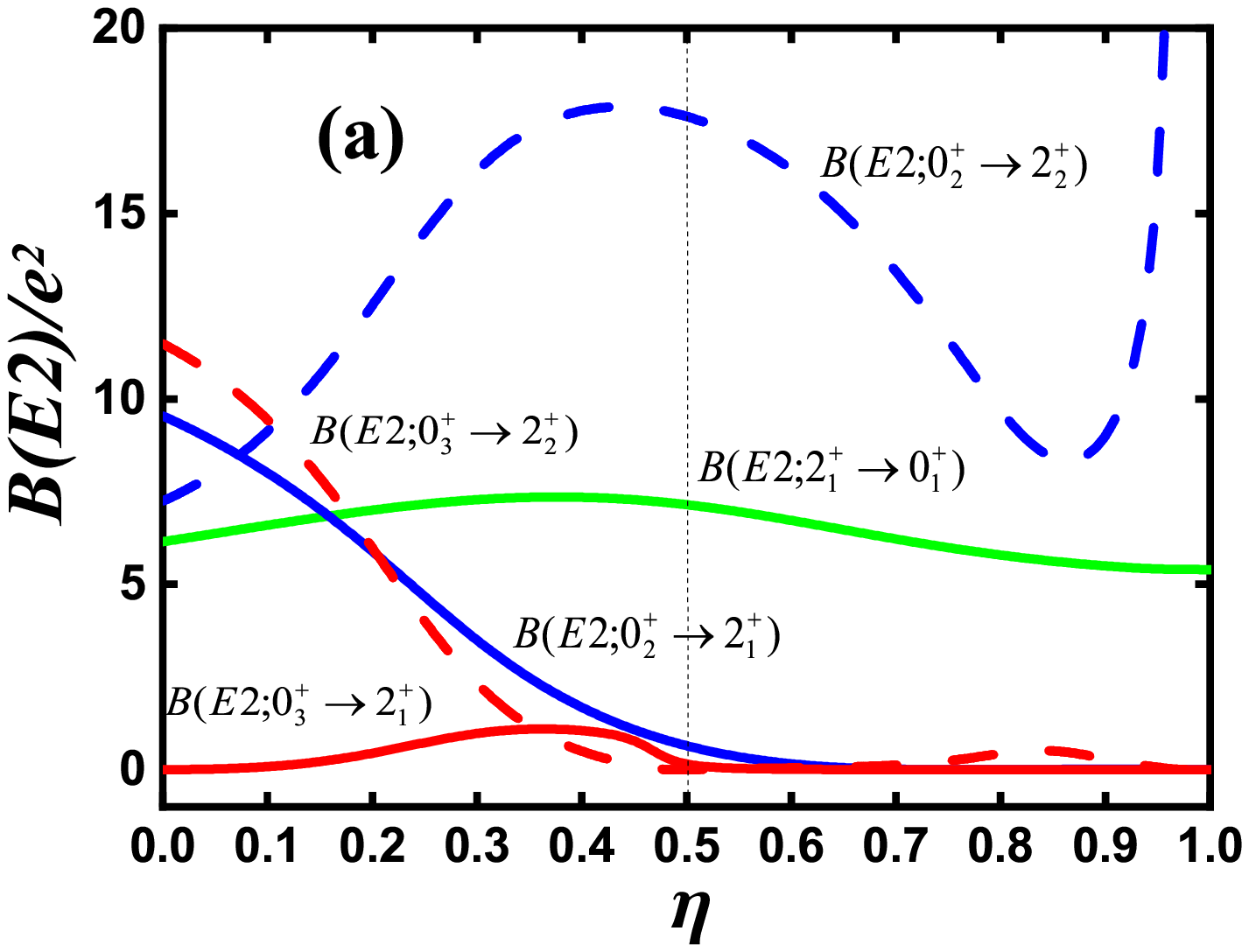}
\includegraphics[scale=0.33]{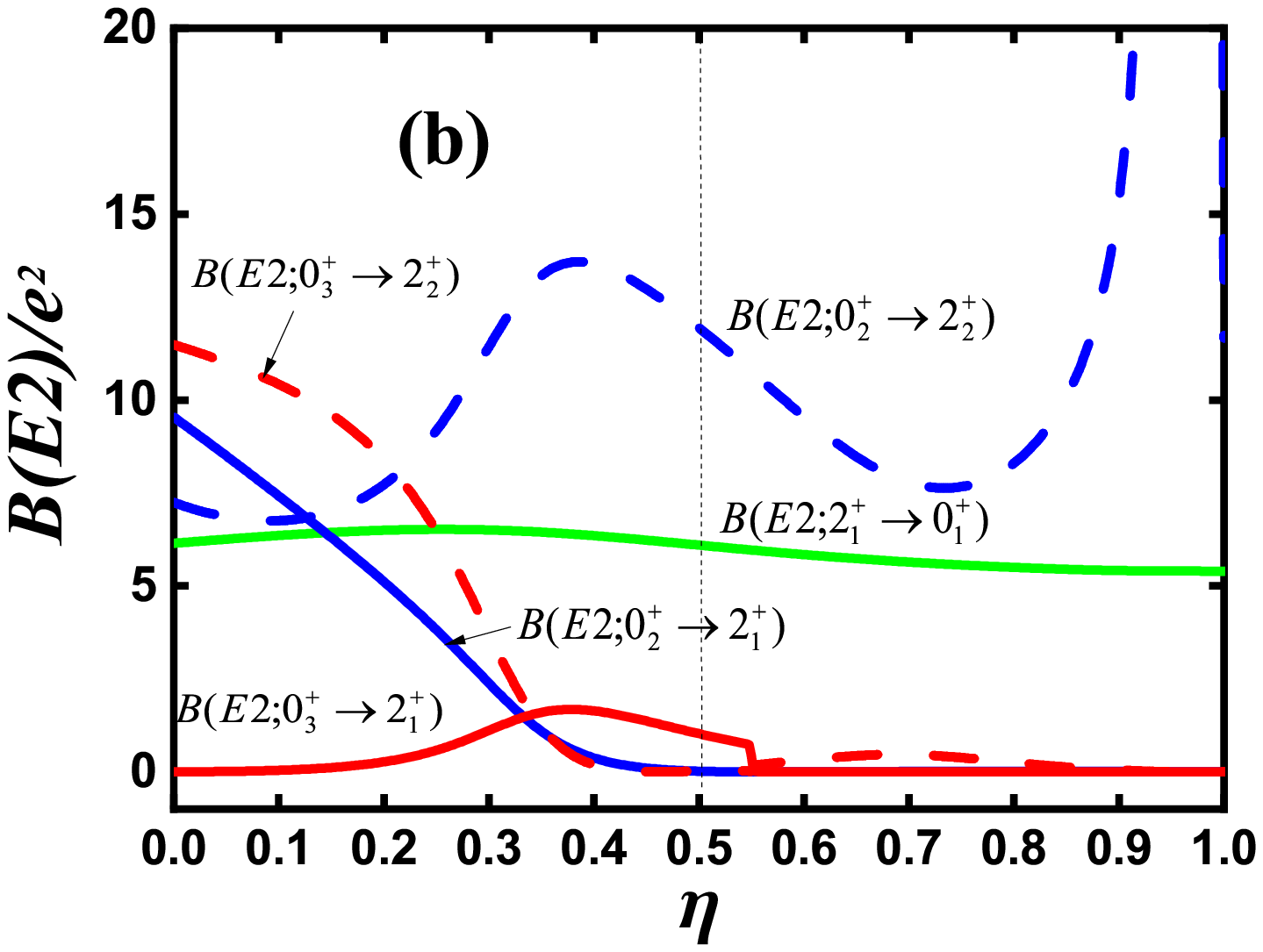}
\caption{The evolution of the $B(E2; 2_{1}^{+}\rightarrow 0_{1}^{+})$ (green real line), $B(E2; 0_{2}^{+}\rightarrow 2_{1}^{+})$ (blue real line), $B(E2; 0_{2}^{+}\rightarrow 2_{2}^{+})$ (blue dashed line), $B(E2; 0_{3}^{+}\rightarrow 2_{1}^{+})$ (red real line), and $B(E2; 0_{3}^{+}\rightarrow 2_{2}^{+})$ (red dashed line) as the function of $\eta$ when $\kappa=1.404$, and (a) $\xi=0$, (b) $\xi=0.05$  for $N=6$.}
\end{figure}

\textbf{In \cite{Wang22}, the normal states in the Cd isotopes are proposed to be a new collective excited mode. The key observation is that there is no the $0_{3}^{+}$ state at the three phonon level while there is a $0_{2}^{+}$ state at the two phonon level. Thus the $0_{3}^{+}$ state is repelled to a higher level and the energy ratio betwen the $0_{3}^{+}$ state and the $0_{2}^{+}$ state is nearly 2.0. Although this large ratio is vital for the new spherical-like spectra, it is impractical for fitting the $^{196}$Pt. For the point $B$, this large ratio still exists, see Fig. 4 at $\xi=0.0$.}  In Ref. \cite{zhou23}, it is found that, the introduction of the fourth-order interaction $\hat{C}_{2}^{2}[\textrm{SU(3)}]$ can reduce the energy difference between the $0_{2}^{+}$ and $0_{3}^{+}$ states, even to zero. Fig. 4 presents the level evolutions of $0_{2}^{+}$ and $0_{3}^{+}$ states when $\xi$ increases from 0 to 0.3 for $\eta=0.5$ and $\kappa=1.404$. It should be noticed that the fourth-order interaction is only a supplementary term here \cite{wang23}, so the value of $\xi$ is small. The distance between the two states reduces first and then gets bigger. The minimum value is around $\xi=0.144$, at which the energy ratio is \textbf{1.188, much smaller than 2.0.}

\begin{figure}[tbh]
\includegraphics[scale=0.33]{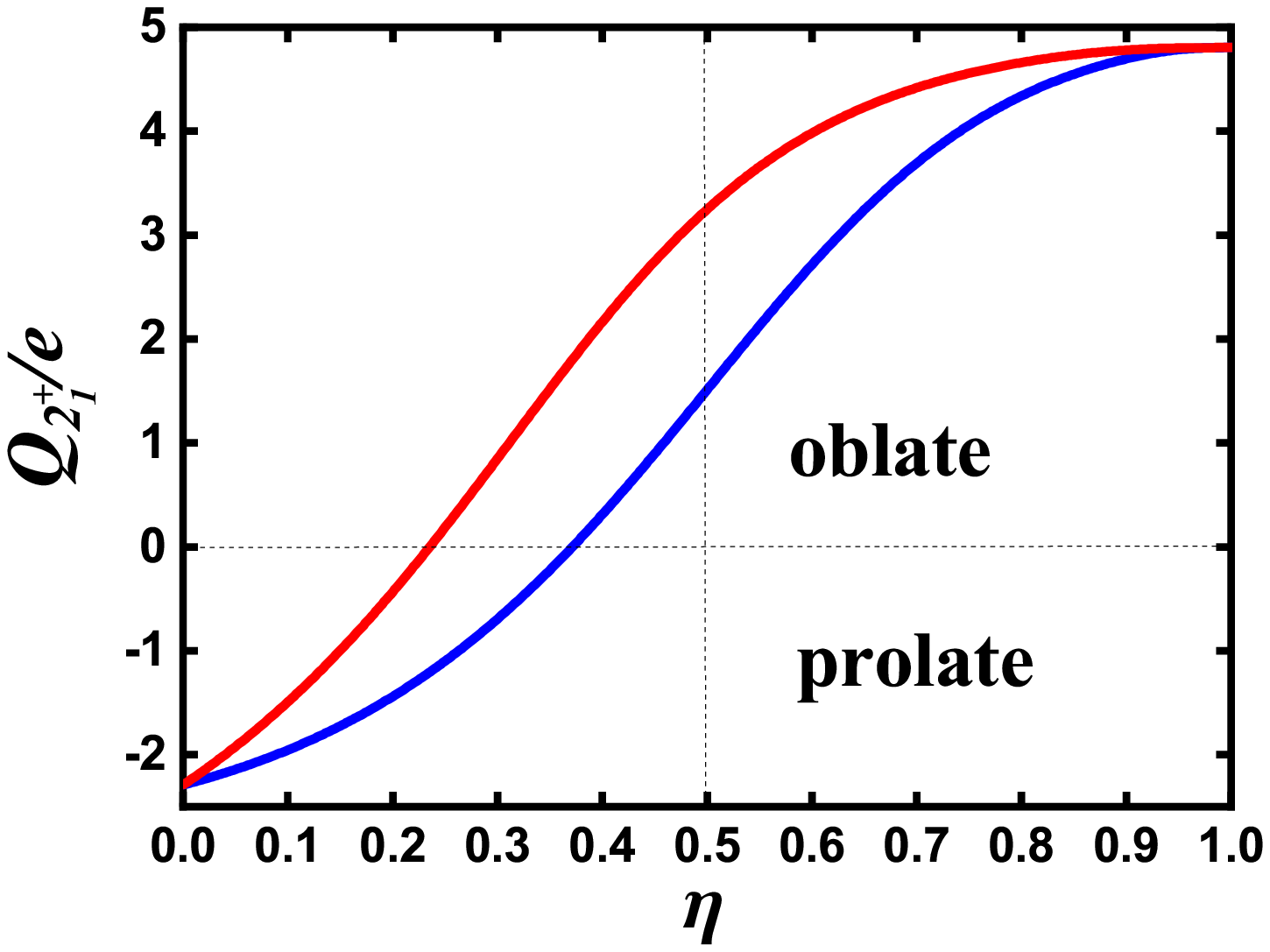}
\caption{The evolution of the quadrupole moment of the $2^{+}_{1}$ state along the blue line in Fig. 1 for $N=6$ when $\xi=0$ (the solid blue line) and $\xi=0.05$ (the solid red line).}
\end{figure}

Fig. 5 (a) presents the partial low-lying level evolutions along the blue line via the point $B$. The degeneracy between the ground band and the $\gamma$-band are somewhat broken, which can be also seen in Fig. 6 (a) for the energy spectra of the point $B$. Compared with \textbf{the top figure in} Fig. 2, the levels of $0_{2}^{+}$ and $2_{3}^{+}$ states move up a little and $6_{1}^{+}$, $4_{2}^{+}$, $3_{1}^{+}$ and $0_{2}^{+}$ are nearly degenerate. Thus the low-lying part of the spectra is similar to that shown in the \textrm{O(6)} limit \cite{Iachello87}. For point $B$, the energies of $0_{2}^{+}$ and $0_{3}^{+}$ states are 1.1637c and 2.1472c in Fig. 5 (a). The ratio of the two states $R'=E_{0_{3}^{+}}/E_{0_{2}^{+}}=1.845$ is much larger than the experimental one 1.236. In this paper, for a better fitting of the quadruple moment, we select $\xi=0.05$. Fig. 5 (b) presents the partial low-lying level evolutions when $\eta$ varies from 0 to 1 for $\kappa=1.404$ and $\xi=0.05$. The most significant change is that the positions of the $0_{2}^{+}$ state increases from 1.16c to 1.47c. The spectra of the position can be seen in Fig. 6 (b), which seems somewhat irregular. But the low-lying part looks very similar to the spectra of the realistic $\gamma$-soft nucleus. For comparison, the bottom figure in Fig. 2 presents the spectra of the point $A$ for $N=6$, which is not shown in \cite{Wang22} and seems different from the two figures in Fig. 6. The low-lying part is very similar to the vibrational spectra of a rigid spherical nucleus. The key difference is that some bands ($0_{4}^{+}$, $2_{8}^{+}$, $0_{6}^{+}$ as bandheads) are greatly elevated. When moving from the point $A$ to the point $B$, then with the adding of fourth-order term, the degeneracies are gradually broken and the familiar $\gamma$-soft feature emerges, while the regularity of the spectra is weakened.

\section{$B(E2)$ values and quadrupole moments for the point $B$}

The $B(E2)$ values are vital for understanding the collective behaviors. In the common experiences of nuclear structure studies, we often expect a definite relationship between the energy spectra and the corresponding $B(E2)$ values, especially in the IBM. However this relation may lead to the wrong conclusions. In the spherical nucleus puzzle \cite{Garrett12,Heyde16,Garrett18}, the energy spectra of the Cd isotopes are similar to the ones of the rigid spherical vibrations, but the $B(E2)$ values are experimentally found to violate the expectations.  Thus new perspectives on the shape evolution from the magic number nucleus to the deformation need to be developed. In the $B(E2)$ anomaly \cite{168Os,166W,172Pt,170Os}, this case becomes more obvious. From the level evolutions of the Pt-Os-W isotopes with neutron number, the energy spectra of $^{172}$Pt, $^{168,170}$Os, $^{166}$W seem normal, but their $B(E2)$ values completely exceed expectations. Thus collective behaviors cannot be determined solely by the energy spectra. Various nuclear spectroscopic methods are needed \cite{Garrett18}.

For understanding $\gamma$-softness, the $B(E2)$ values are also necessary. Especially when the new $\gamma$-softness is provided \cite{Wang22}, how to distinguish the different $\gamma$ softness becomes more and more important in the description of the realistic nuclei properties. The $E2$ operator is defined as
\begin{equation}
\hat{T}(E2)=e\hat{Q},
\end{equation}
where $e$ is the boson effective charge. The evolution of the $B(E2; 2_{1}^{+}\rightarrow 0_{1}^{+})$, $B(E2; 0_{2}^{+}\rightarrow 2_{1}^{+})$, $B(E2; 0_{2}^{+}\rightarrow 2_{2}^{+})$, $B(E2; 0_{3}^{+}\rightarrow 2_{1}^{+})$, and $B(E2; 0_{3}^{+}\rightarrow 2_{2}^{+})$ values are plotted along the blue line in Fig. 1 for $N=6$. In Fig. 7 (a) the $B(E2; 2_{1}^{+}\rightarrow 0_{1}^{+})$ value is nearly the same. For $\eta=1.0$, it describes an oblate shape \cite{Zhang12}, thus the $B(E2; 2_{1}^{+}\rightarrow 0_{1}^{+})$ value is suppressed. With the increasing of $\eta$, the values of $B(E2; 0_{2}^{+}\rightarrow 2_{1}^{+})$ and $B(E2; 0_{3}^{+}\rightarrow 2_{2}^{+})$ get reduced while the ones of the $B(E2; 0_{2}^{+}\rightarrow 2_{2}^{+})$ becomes larger. The trends are similar to the ones along the green line with degeneracy and the $\gamma$-softness can emerge. When the fourth-order interaction is introduced in Fig. 7 (b), at $\eta=0.5$, the value of $B(E2; 0_{2}^{+}\rightarrow 2_{2}^{+})$ can be reduced.

Fig. 8 shows the quadrupole moments of the $2^{+}_{1}$ state for $\xi=0$ (the solid blue line) and $\xi=0.05$ (the solid red line).  It is shown that, for the blue line, when $\eta\geq 0.372$, the value becomes positive, which means an oblate deformation. For the red line, it bends to the oblate side.  \textbf{It should be noticed that the $E2$ operator is suitable here. For the U(5) limit, the quadrupole operator is  $[d^{\dag}\times \tilde{d}]^{(2)}$, with which the quadrupole moment of the $2^{+}_{1}$ state in the U(5) limit is negative (not zero) \cite{Iachello87}. If the quadrupole operator is taken as that of the O(6) limit, namely $\hat{Q}_{0}=[d^{\dag}\times\tilde{s}+s^{\dag}\times \tilde{d}]^{(2)}$, then the quadrupole moment of the $2^{+}_{1}$ state in the U(5) limit is just zero. Thus, if $\hat{Q}=[d^{\dag}\times\tilde{s}+s^{\dag}\times \tilde{d}]^{(2)}-\frac{\sqrt{7}}{2}[d^{\dag}\times \tilde{d}]^{(2)}$ is used, in the SU(3) limit, it is the $\hat{Q}$ while in the U(5) limit it is just the $[d^{\dag}\times \tilde{d}]^{(2)}$. This can describe the real evolution. In the actual fitting, the key is that the effective charge will be different.}

\section{Theoretical fitting of $^{196}$Pt}

\begin{figure}[tbh]
\includegraphics[scale=0.29]{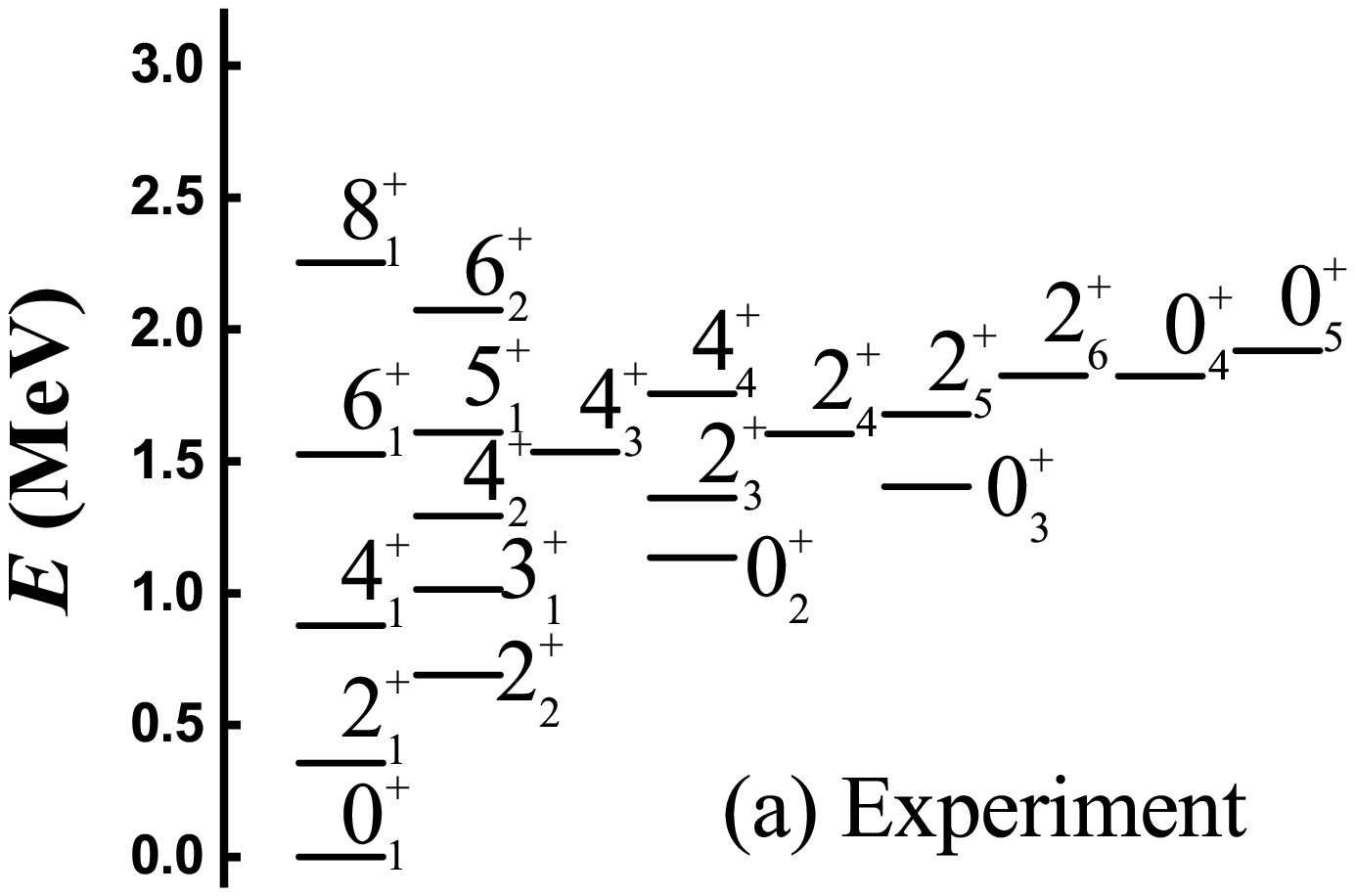}
\includegraphics[scale=0.29]{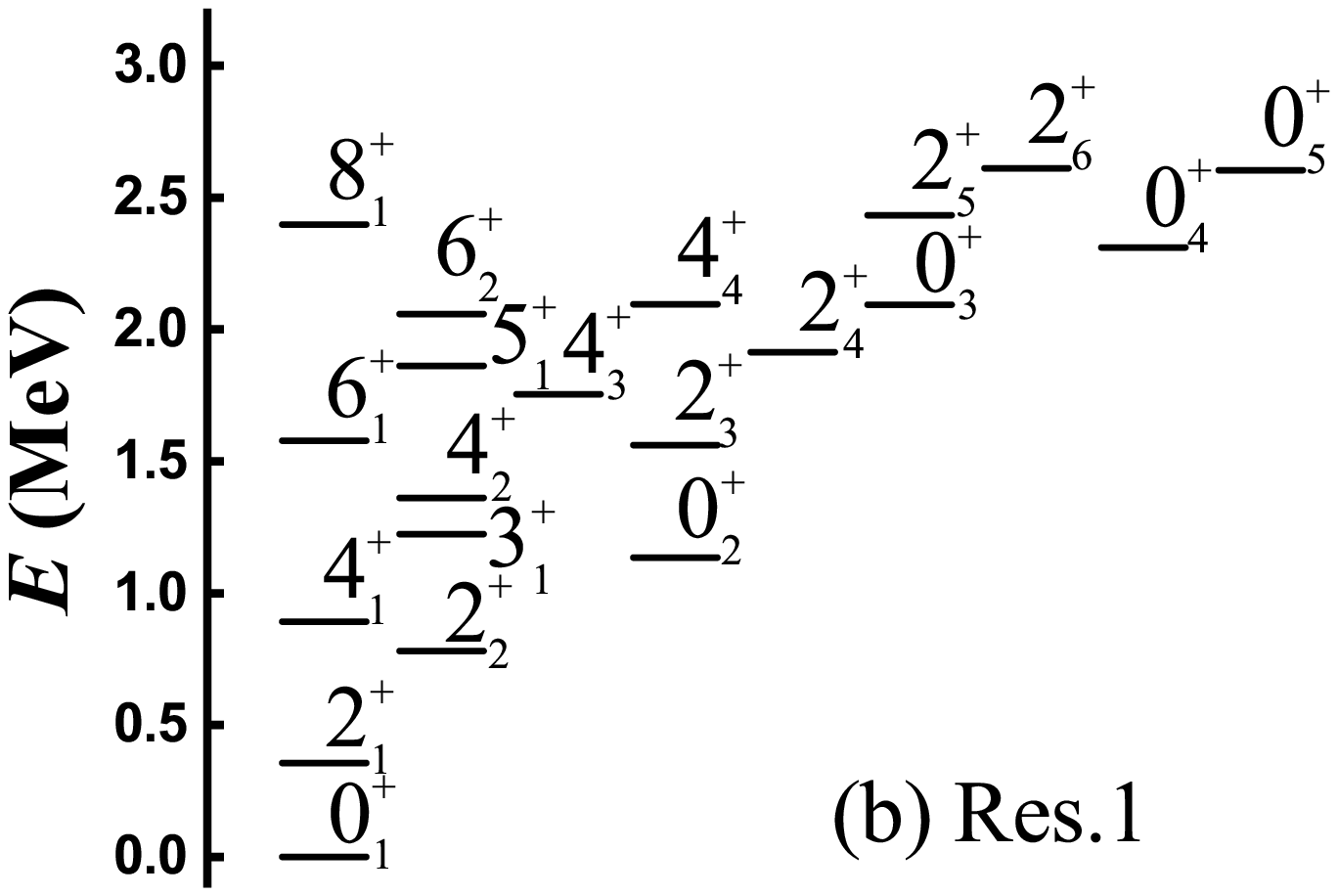}
\includegraphics[scale=0.29]{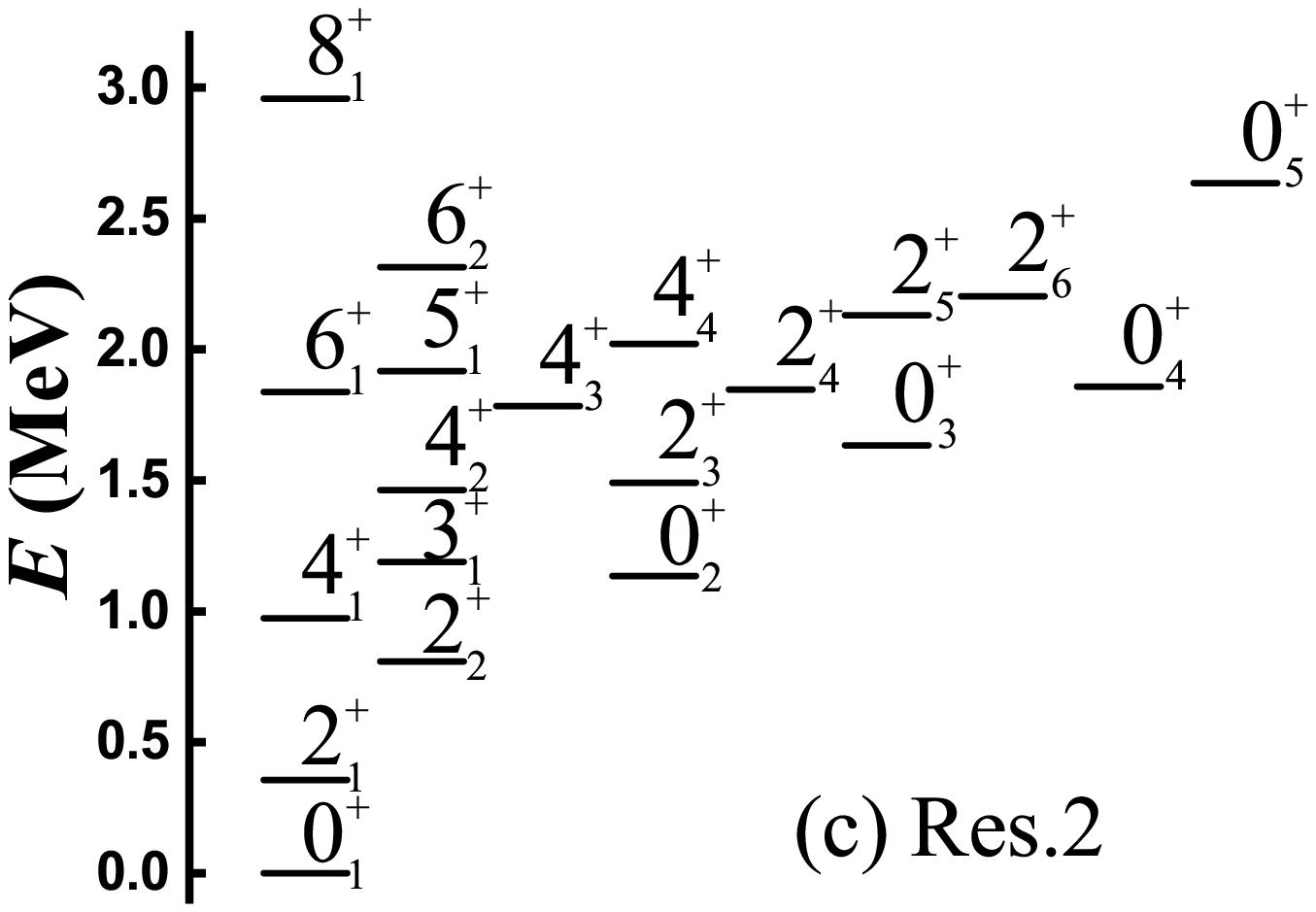}
\caption{Energy spectra of (a) $^{196}$Pt, and of the Hamiltonian in Eq. (3) at the point $B$ (b) and when adding the fourth-order interaction when $\xi=0.05$ for $N=6$ (c).}
\end{figure}

Without considering other higher-order interactions in the \textrm{SU(3)} limit, the properties of the point $B$ or adding the fourth-order interaction are used to fit the structure of $^{196}$Pt. Although the precision needs to be improved, the fitting results seem excellent. For $\xi=0$ at point $B$, the overall energy parameter $c$ in $\hat{H}$ is 0.9753 MeV to make the energy value of the $0_{2}^{+}$ state equal to the experimental one. The $L^{2}$ term is also added to fit the $2_{1}^{+}$ state, which is 0.00803 MeV. The theoretical spectra of point $B$ shown in Fig. 9 (b) are compared with the experimental data shown in Fig. 9 (a). The theory and the experiment correspond well qualitatively, and we can see that the position relationships of each energy level are also consistent. The rotational-like $\gamma$-band in $^{196}$Pt is an interesting problem, and the theoretical spectra have similar structures. $0_{4}^{+}$ and $0_{5}^{+}$ states also fit well. The main drawback is that the $0_{3}^{+}$, $2_{5}^{+}$ and $2_{6}^{+}$ are somewhat higher, that is, the energy difference between the $0_{3}^{+}$ and $0_{2}^{+}$ states is somewhat larger than the experimental result. This is the typical feature of the new $\gamma$-soft-like rotational mode \cite{Wang22}. For $\xi=0.05$, better fitting results can be obtained, where the characteristics of the $\gamma$ band are consistent with the actual situation and the energies of the $0_{3}^{+}$ and $0_{4}^{+}$ states are also reduced. For reducing the energies of the higher-levels, Pan \emph{et al.} presented a new method to provide an excellent fitting result for $^{194}$Pt \cite{Pan18}, which may be used to improve the fitting precision in the SU3-IBM.

\begin{figure}[tbh]
\includegraphics[scale=0.33]{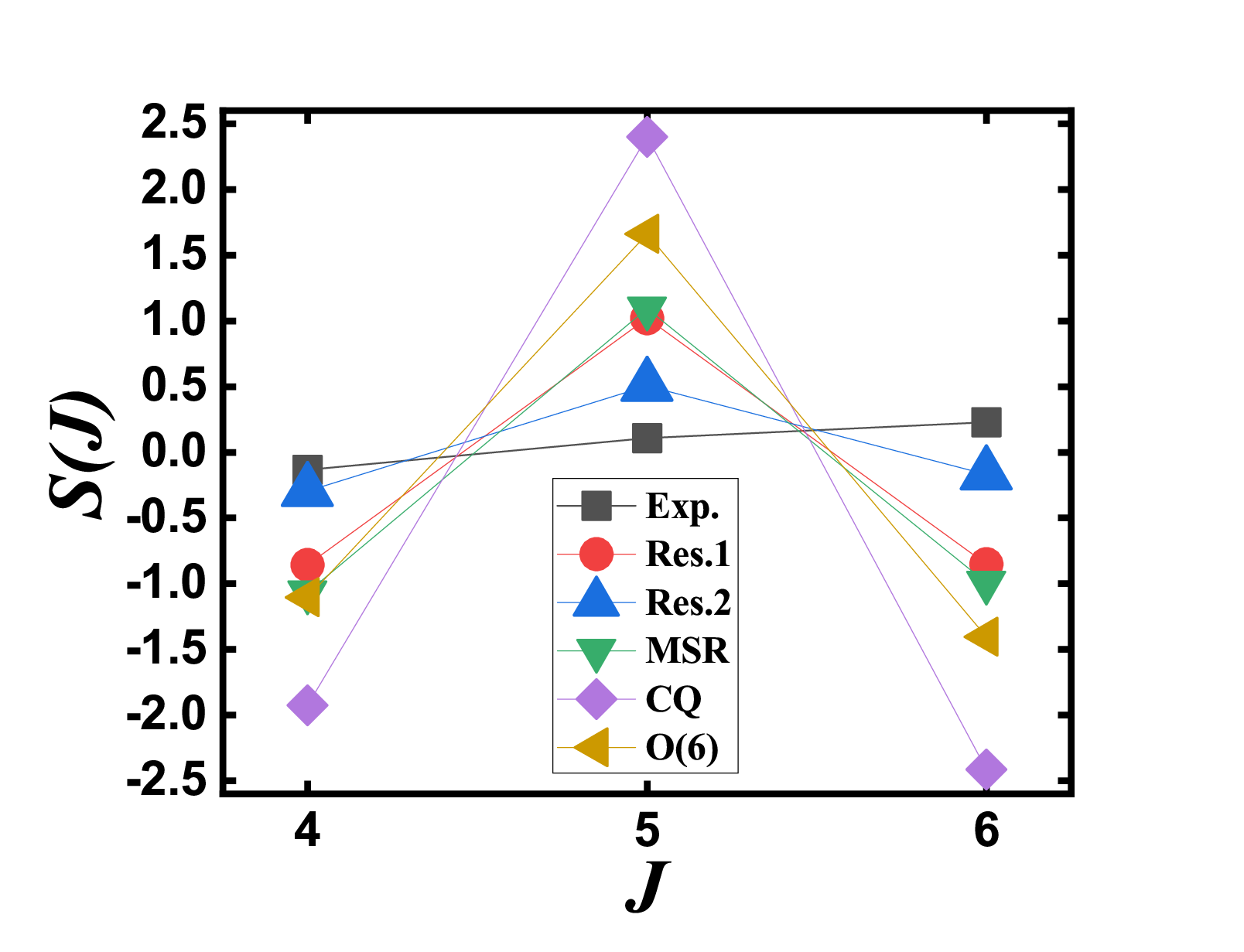}
\caption{The staggering parameter $S(J)$ for $^{196}$Pt and various models.}
\end{figure}

\begin{table}[tbh]
\caption{\label{table:expee}  Absolute $B(E2)$ values in W.u. for $E2$ transitions from the low-lying states in $^{196}$Pt, Res.1 for the cousin point $B$, Res.2 for adding the fourth-order interaction when $\xi=0.05$, the PDS model and the MSR model with effective charge $e=2.385$ (W.u.)$^{1/2}$ for Res.1 and $e=2.623$ (W.u.)$^{1/2}$ for Res.2. $^{a}$From Ref.\cite{sheet07}, $^{b}$From Ref.\cite{Leviatan09}, $^{c}$From Ref.\cite{Li22}, $^{d}$From Ref.\cite{Li22}.}
\setlength{\tabcolsep}{0.1mm}{
\begin{tabular}{cccccccc}
\hline
\hline
$L_{i}  ~~L_{f}    $ &Exp.$^{a}$                \ &Res.1    \ &Res.2           \ &PDS$^{b}$    \ &MSR$^{c}$    \ &CQ$^{d}$    \\
 \hline
$2_1^+  ~~ 0_1^+   $ & 40.60(20)                \ & 40.6       \ & 40.6            \ & 40.6       \ &40.6      \  & 40.6 \\
$2_2^+  ~~ 2_1^+   $ & 54(+11 -12)              \ & 60.3       \ & 43.2            \ & 53.0       \ &45.3      \ & 50.2 \\
$~~~~   ~~  0_1^+  $ & $<7.8\times 10^{-6}$     \ & 2.9        \ & 5.05            \ & 0.27       \ &1.4       \ & 0.31   \\
$4_1^+  ~~  2_1^+  $ & 60.0(9)                  \ & 52.6       \ & 51.4            \ & 53.0       \ &53.6      \ & 53.6 \\
$0_2^+  ~~ 2_1^+   $ & 2.8(15)                  \ & 3.6        \ & 0.122           \ & 0.44       \ &3.1       \ &  0.55 \\
$~~~~   ~~ 2_2^+   $ & 18(10)                   \ & 100.0      \ & 79.5            \ & 54.1       \ &69.6      \ &  51.9 \\
$6_1^+  ~~ 4_1^+   $ & 73(+4 -73)               \ & 52.1       \ & 45.3            \ & 54.1       \ &54.7      \ &  54.7 \\
$4_2^+  ~~ 4_1^+   $ & 17(6)                    \ & 34.4       \ & 27.4            \ & 25.8       \ &18.1      \ & 23.8\\
$~~~~   ~~ 2_1^+   $ & 0.56(+12 -17)            \ & 0.65       \ & 1.34            \ & 0.14       \ &0.69      \  &   0.002    \\
$~~~~   ~~  2_2^+  $ & 29(+6 -29)               \ & 38.0       \ & 31.2            \ & 28.3       \ &28.8      \  &  28.3    \\
$2_3^+  ~~ 4_1^+   $ & 0.13(12)                 \ & 2.34       \ & 0.23           \ & 0.059      \ &0.60       \  &  0.37   \\
$~~~~   ~~ 2_2^+   $ & 0.26(23)                 \ & 3.93       \ & 3.47           \ & 0.19       \ &0.27       \ &  0.09  \\
$~~~~   ~~  0_1^+  $ & 0.0025(24)               \ & 0.025      \ & 0.02           \ & 0          \ &0.064      \  &   0.033   \\
$~~~~   ~~  0_2^+  $ & 5(5)                     \ & 38.3       \ & 36.5          \ & 17.6       \ &0.003       \  &  11.08    \\
$0_3^+  ~~  2_1^+  $ & $<5.0$                   \ & 0.88       \ & 6.82            \ & 0          \ &1.05      \  &  0.38  \\
$~~~~   ~~  2_2^+  $ & $<0.41$                  \ & 0.015      \ & 0.114        \ & 0.41       \ &0.009        \  &  1.22   \\
$6_2^+  ~~  6_1^+  $ & 16(5)                    \ & 23.2       \ & 17.7          \ & 15.3       \ &9.5         \  & 0.02  \\
$~~~~   ~~  4_1^+  $ & 0.48(14)                 \ & 0.29       \ & 0.77          \ & 0.12       \ &0.50        \ &  0.0003 \\
$~~~~   ~~  4_2^+  $ & 49(13)                   \ & 42.1       \ & 31.9          \ & 32.7       \ &32.9        \   &  0.008   \\
$8_1^+  ~~  6_1^+  $ & 78(+10 -78)              \ & 49.1       \ & 29.3        \ &            \ &47.6          \  & 48.5 \\
\hline
\hline
\end{tabular}}
\end{table}

\begin{table}[tbh]
\caption{\label{table:expee} $R'$ and $R$ in $^{196}$Pt and various models. $^{a}$From Ref.\cite{sheet07}, $^{b}$From Ref.\cite{Leviatan09}, $^{c}$From Ref.\cite{Li22}.}
\setlength{\tabcolsep}{0.2mm}{
\begin{tabular}{cccccccc}
\hline
\hline
 &Exp.$^{a}$                \ &Res.1       \ &Res.2        \ &MSR$^{c}$  \ &CQ$^{c}$    \ &O(6)$^{b}$  \ &PDS$^{b}$  \\
 \hline
$R'$& 1.236                  \ & 1.845       \ & 1.440             \ & 1.131 \ &1.281  \ &1.543  \ &1.786 \\
$R $& -0.701                    \ & -0.896       \ & -0.864          \ & -0.192 \ &-0.286  \ &-0.289  \ &-0.629 \\
\hline
\hline
\end{tabular}}
\end{table}

\begin{table}[tbh]
\caption{\label{table:expee}  Quadrupole moments in eb and average variance $\Delta Q$ for some low-lying states in $^{196}$Pt and various models. $^{a}$From Ref.\cite{sheet07}, $^{b}$From Ref.\cite{Li22}. }
\setlength{\tabcolsep}{0.6mm}{
\begin{tabular}{ccccccc}
\hline
\hline
$                  $ &Exp.$^{a}$                \ &Res.1       \ &Res.2        \ &MSR$^{b}$  \ &CQ$^{b}$  \\
 \hline
$Q(2_{1}^{+})      $ & +0.62                    \ & +0.30        \ & +0.70             \ &+0.50   \ &+0.40   \\
$Q(2_{2}^{+})      $ & -0.39                    \ & -0.48        \ & -0.81           \ &-0.47   \ &-0.37    \\
$Q(4_{1}^{+})      $ & +1.03                    \ & +0.036     \ & +0.63              \ &+0.66     \ &+0.40      \\
$Q(6_{1}^{+})      $ & -0.18                    \ & -0.42       \ &+0.35          \ &+0.76  \ &+0.38    \\
$\Delta Q      $ &                    \ & 0.54      \ &0.39        \ &0.51 \ &0.44  \\
\hline
\hline
 \end{tabular}}
\end{table}

Table I lists the $B(E2)$ values of some low-lying states in $^{196}$Pt, the point $B$ (Res.1), adding the fourth-order term (Res.2), the \textrm{O(6)} partial dynamical symmetry model (PDS) \cite{Leviatan09}, the modified soft-rotor model (MSR) \cite{Li22} \textbf{and the consistent-$Q$ (CQ) \cite{Li22}}. These \textbf{previous} three models are all related to higher-order interactions in IBM. In the \textrm{O(6)} partial dynamical symmetry, one three-body interactions that is partially solvable in \textrm{O(6)} symmetry can be constructed, which can mix the $\Sigma =4$ and $\Sigma=2$, but it does not change the case $\Sigma =6$ for $N=6$ ($\Sigma$ is the \textrm{O(6)} label). In the modified soft-rotor model, the higher-order interactions are used to fit the Pt isotopes, which is inspired by the \textrm{O(6)} higher-order symmetry description of $^{194}$Pt \cite{Pan18}. \textbf{For comparison, the CQ results are also listed here.} For Res.1, this $\gamma$-soft-like description of the point $B$ can show a good consistency with the experimental data qualitatively \cite{sheet07}. From the overall fitting results, it looks like somewhat worse than the other two theories, but the result is still good considering that the parameters of the point $B$ are not adjustable. When the fourth-order interactions are introduced for $\xi=0.05$, the fit can be greatly improved.

To better substantiate this conclusion, a quantitative analysis is mandatory. The first quantity that we might study is the staggering parameter $S(J)$ in $\gamma$ band energies \cite{Casten91,Casten07} defined as
\begin{equation}
S(J)=\frac{(E_{J}-E_{J-1})-(E_{J-1}-E_{J-2})}{E_{2_{1}^{+}}},
\end{equation}
which quantifies how adjacent levels within a $\gamma$ band are grouped. Fig. 10 presents the $S(J)$ for $J=4,5,6$ in experimental data, Res.1 and Res.2 in the SU3-IBM, MSR \cite{Li22}, CQ \cite{Li22} and \textrm{O(6)} symmetry \cite{Leviatan09}. Black squares are the experimental results. CQ shows the typical $\gamma$-soft feature of strong staggering. \textrm{O(6)}, MSR and Res.1 display similar trends, while Res.2 gives the best fitting.

\begin{figure}[tbh]
\includegraphics[scale=0.29]{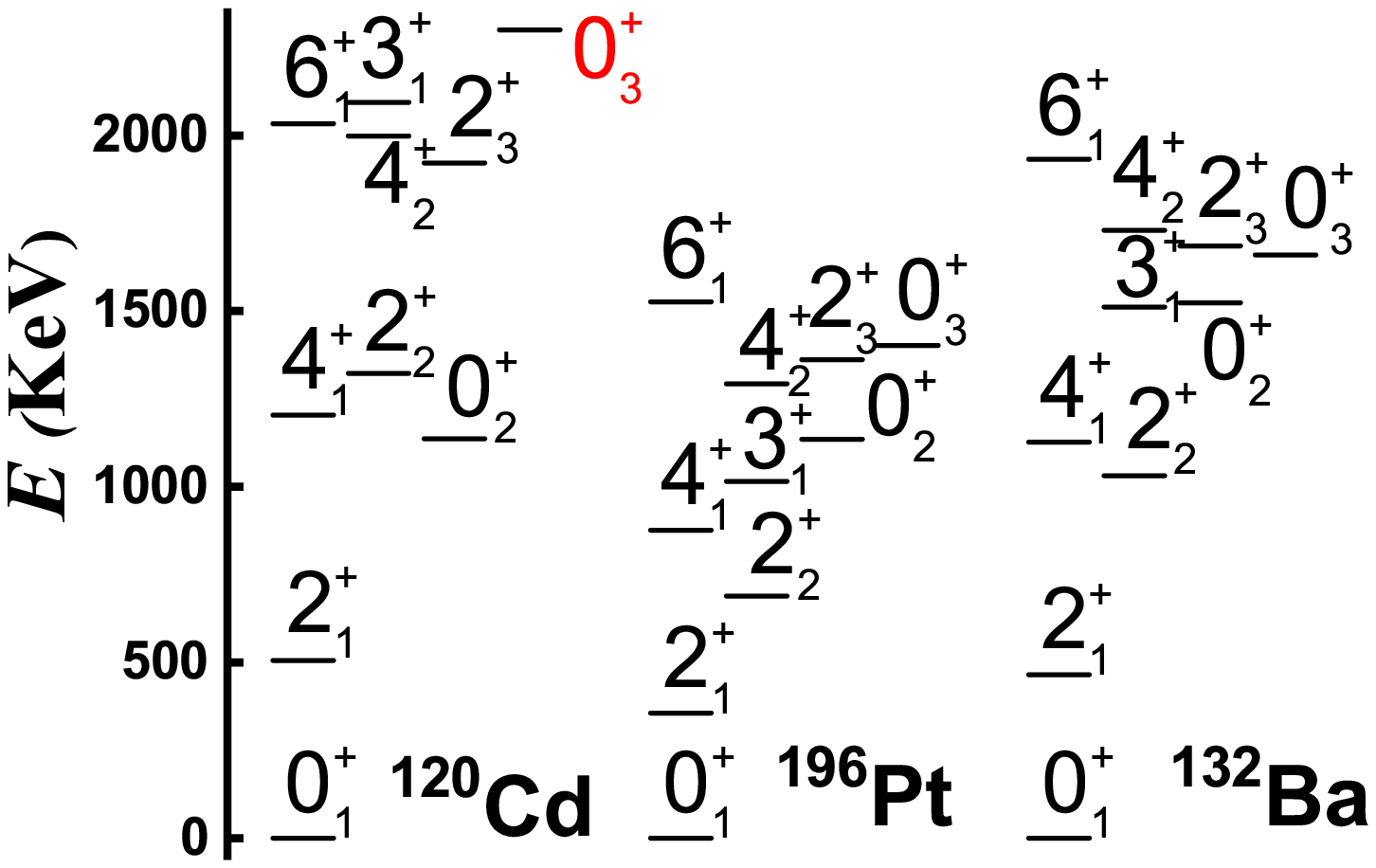}
\caption{Partial energy spectra in $^{120}$Cd normal states, $^{196}$Pt and $^{132}$Ba \cite{ensdf}. The energy of the $0_{3}^{+}$ state in the $^{120}$Cd normal states is predicted by the authors.}
\end{figure}

The second is on the positions of the $0_{2}^{+}$, $0_{3}^{+}$ and $0_{4}^{+}$ states in the spectra. $R'=E_{0_{3}^{+}}/E_{0_{2}^{+}}$ is the energy ratio between the $0_{2}^{+}$ and $0_{3}^{+}$ states. $R=E_{0_{4}^{+}}/E_{0_{3}^{+}}-2$ is used in \cite{Leviatan09}. Table II presents the $R'$ and $R$ in experimental data, Res.1, Res.2, MSR \cite{Li22}, CQ \cite{Li22}, O(6) symmetry \cite{Leviatan09} and PDS \cite{Leviatan09} . For $R'$, Res.2, MSR and CQ can offer reasonable results. For $R$, \textbf{Res.1}, Res.2 and PDS can give the best results that are consistent with the experiment data. In \cite{Leviatan09}, although the introduction of higher-order interaction can fit the $R$ well, it increases $R'$. Res.2 is the only theory that makes both values more consistent. What needs to be mentioned is that the $0_{4}^{+}$ sate in $^{194}$Pt is an intruder state \cite{Pan18}, thus the experimental $0_{5}^{+}$ (or $0_{4}^{+}$) state in $^{196}$Pt may be also an intruder state.

The experimental data about the quadrupole moments are rare, which are very sensitive to specific nuclear structure models. Table III presents the quadrupole moments of some low-lying states in $^{196}$Pt, which contain the results of Res.1, Res.2, the MSR model \cite{Li22} and CQ \cite{Li22}. The average variance of the fit $\Delta Q$ are calculated. Our results are more consistent with the experimental data. The quadrupole moment of these states in the $O(6)$ limit are all zero with $\Delta Q=0.64$. The quadrupole moments can be a very useful indicator for judging the success of different nuclear models.

The results of these three quantitative calculations favour the \textbf{description} in terms of SU3-IBM over previous theories.

\section{Some discussions about the emerging $\gamma$-softness}

\textbf{In the traditional view of nuclear structure evolution, the collective excitation begins with the phonon excitation mode of the spherical nucleus, just away from the magic number nucleus (the boson number is small). When the boson number increases further, the deformation appears, and then a large deformation gradually emerges, such as the prolate ellipsoid, which is typically characterized by its rotational spectra. Eventually, the $\gamma$-soft spectra appear when approaching the next magic nucleus. An authoritative review on this shape phase transitions appeared in 2010 \cite{Casten10}, but today these ideas have faced serious challenges.}
The spherical nucleus puzzle \cite{Garrett08,Garrett10,Garrett12,Batchelder12,Heyde16,Heyde16,Heyde16,Garrett18,Garrett19,Garrett20} and $B(E2)$ anomaly \cite{168Os,166W,172Pt,170Os} dramatically change our view of \textbf{the origin of collectivity} in nuclear structure and shape evolutions. \textbf{If the simple spherical shape does not exist, what collective excitation is it? It is also noticed that even for magic nuclei the deformation might already be present \cite{Garrett19,Otsuka18}.} These two peculiar phenomena occur in the collective modes of nuclei with boson number $N<10$. This means that the emergence of collective motions in realistic nuclei will be a more complex problem. Holding an obstinate simple view of collective motions in nuclei seems very undesirable. Further experimental information on the two puzzles may lead us to get a more accurate description on nuclear structures. \textbf{In the SU3-IBM, the spherical shape is replaced by a spherical-like quadrupole deformation, which challenges the traditional view.}

Even for typical large-deformation nuclei as $^{154}$Sm and $^{166}$Er, new studies have found that the structure of $^{154}$Sm shows the coexistence of prolate and triaxial shapes, while the deformed shape with a strong triaxial instability is demonstrated on $^{166}$Er, which has shown to be related with self-organization mechanism \cite{Otsuka19,Otsuka22}. \textbf{This perspective seems very attractive. In the traditional view, $^{154}$Sm and $^{166}$Er are two typical nuclei having prolate shape with $\beta/\gamma$ vibrational excitations for the rotational spectra exists. However, recently, some nuclear structure models have been used to investigate the nature of $^{166}$Er \cite{Otsuka19,Otsuka21,Frauendorf24}, and it was found that the ground state of this largely-deformed nucleus really has triaxial deformation with $\gamma=9^{\circ}$. Otsuka \textit{et al.} argue that the triaxial shapes dominates in the heavy nuclei \cite{Otsuka24}. These studies questioned the prolate predominance and placed the triaxial shapes in a much more important position.}

\textbf{In our study, on the prolate-oblate asymmetic shape phase transition \cite{wang23}, only the SU(3) second-order (prolate shape) and third-order (oblate shape)  Casimir operators $\hat{C}_{2}[\textrm{SU(3)}]$, $\hat{C}_{3}[\textrm{SU(3)}]$ are considered. In the SU(3) limit, the prolate shape is described by the SU(3) irrep (2$N$,0) while the oblate shape is (0,$N$). However these are not enough. In this paper, taking $^{196}$Pt as an example, the square of the SU(3) second-order operator  $\hat{C}_{2}^{2}[\textrm{SU(3)}]$ is necessary. In \cite{zhou23}, it showed that this fourth-order interaction is vital for the triaxial shape. A new study shows that $^{196}$Pt may be related with the SU(3) irrep (4,4) \cite{zhou24}. From the perspective of Otsuka \textit{et al.} \cite{Otsuka19,Otsuka21,Otsuka24}, this interaction may be also important for other nuclei, even the prolate nuclei $^{180}$Hf. The discussion of this shae phase transition needs further studies, and SU3-IBM will also be used to study the triaxiality of these large deformation nuclei. For $^{166}$Er, if $\gamma=9^{\circ}$, the ground state has the SU(3) irrep (22,4), which can be easily treated by SU3-IBM. It should be noticed that every quadrupole deformation is treated equally in SU3-IBM, which is different from previous models with SU(3) symmetry.}

\textbf{Besides the spherical nuclei and the prolate nuclei, the O(6) description of the $\gamma$-soft nuclei is also questioned by us in Ref. \cite{wang23,wangtao}.} This article can be seen as an extension of the previous works \cite{Wang22,wang23}. In these papers, the SU3-IBM is proposed, which is inspired by the interesting findings in Ref. \cite{Fortunato11,Zhang12}. When only the \textrm{U(5)} limit and the \textrm{SU(3)} limit are concerned, $\gamma$-soft-like rotational mode can also occur as an emergent phenomenon if the \textrm{SU(3)} higher-order interactions are introduced into the common formalism. This is very different from the $\gamma$-softness related to the \textrm{O(6)} limit. In the traditional IBM, \textrm{O(6)} symmetry is exactly solvable, and $\gamma$-unstable spectra can be expected. Especially the SU3-IBM can describe the $B(E2)$ anomaly, which make the model particularly useful. \textbf{If triaxiality dominates the heavy nuclei, it can further support the SU3-IBM.}

\textbf{{ Since the inception of the $\gamma$-independent solution in the geometric model \cite{Jean56}, this has always been regarded as one of the paradigms in nuclear structure, alongside the spherical and axially-deformed cases. The beahviour of the potential in the $\gamma$ variable can range from the extreme case of $\gamma-$independent potential to the  $\gamma-$rigid case, passing through the intermediate case of $\gamma-$soft. In the soft cases (and up to complete $\gamma-$independence) the fluctuations of triaxiality might be very large, which makes the shape of a $\gamma-$soft much more difficult to understand with repsect to the spherical and axial cases. The ideal O(6) symmetry implies complete $\gamma$-instability \cite{Iachello87}, and it is quite unrealistic to think that nuclei might realize it, even only approximately. The problem comes}
from the $\overline{\textrm{SU(3)}}$ symmetry in Hamiltonian (1), and the O(6) symmetry point is regarded as the critical point between the proalte shape and the oblate shape \cite{Werner01,Jolie03}. $\overline{\textrm{SU(3)}}$ symmetry has the same spectra as the SU(3) symmetry \cite{Wang08}. However this spectral symmetry can not be found in realistic nuclei, which presents many constraints on the shape evolution from the prolate shape to the oblate shape. In \cite{wang23}, it was found that this shape evolution is asymmetric, so the $\overline{\textrm{SU(3)}}$ symmetry cannot be { realized and also the O(6) symmetry, albeit  pretty from the mathematical perspective, cannot be easily attained in realistic nuclei, that can be $\gamma$-soft, but not fully $\gamma$-unstable. Of course, this needs further evidences.} }

\textbf{In the SU3-IBM, the $\gamma$-softness results from the combinations of the $d$ boson number operator and the rigid quadrupole defomations described by the SU(3) interactions. Thus it is an emergent phenomenon, and different from the one with O(6) symmetry.}

Fig. 11 presents partial energy spectra of $^{120}$Cd, $^{196}$Pt and $^{132}$Ba, for their boson number are all $N=6$. Ref. \cite{Batchelder12} obtains the spectra of $^{120}$Cd , in which the $0_{3}^{+}$ state at the three-phonon level is absent. The energy of the $0_{3}^{+}$ state of $^{120}$Cd is predicted by our theory, which is around 2300 KeV, or even higher. Thus this is really a new $\gamma$-soft rotational mode \cite{Garrett10,Garrett12,Garrett18}, despite it looks much like the rigid spherical vibrational excitation mode. If we consider that the IBM can truly describe the collective excitations in realistic nuclei, the only way to describe the new $\gamma$-soft behaviors is to introduce the \textrm{SU(3)} higher-order interactions. The theory works better than expected \cite{Wang22,Wang20,Zhang22}.

$^{196}$Pt and $^{132}$Ba are two typical traditional $\gamma$-soft nuclei. In the IBM, this kind of $\gamma$-softness is related to the \textrm{O(6)} symmetry, or \textrm{O(5)} symmetry, such as \textrm{E(5)} critical point description \cite{Iachello00,Zhangyu22}. In fact, this case is not related to the deformation $\gamma$ parameter. The SU3-IBM follows such a principle that triaxiality arises from the competition between the prolate shape and the oblate shape. In this paper, we show that the new emerging $\gamma$-softness can resemble the traditional $\gamma$-soft spectra and their $B(E2)$ behaviors. When other \textrm{SU(3)} higher-order interactions are introduced, the fitting of spectra appears excellent. This will be further discussed in following papers.

\textbf{This principle requires some additional clarifications. The sign of the electric quadrupole moment of the $2_{1}^{+}$ state in a realistic nucleus provides a robust characteristic of the shape. If the sign is negative, it is prolate shape or prolate-biased triaxiality while if positive, it is oblate shape or oblate-biased triaxiality.
For realistic nuclei, it can be found that the sign can change from negative to positive, and triaxiality must plays a key role for understanding the shape revolution \cite{wang23}. The SU3-IBM is particularly suitable to discuss the evolution of the quadrupole deformations for these shapes can be explicitly and directly given by the SU(3) irrep $(\lambda, \mu)$. In the SU3-IBM, the evolution from the prolate shape to the oblate shape can easily include the triaxial shapes, and in fact there are many evolutional paths including triaxiality. Even for the SU(3) degenerate point, it contains various possible quadrupole deformations \cite{Wang22,Zhang12}. When $\hat{C}_{2}^{2}[\textrm{SU(3)}]$ is introduced, the ground state can have a certain rigid triaxial shape in the SU(3) limit \cite{zhou23}. It should be noticed that $\hat{C}_{2}^{2}[\textrm{SU(3)}]$ can not describe the triaxiality alone, and should be used with the $\hat{C}_{2}[\textrm{SU(3)}]$, $\hat{C}_{3}[\textrm{SU(3)}]$ interactions. $\hat{C}_{2}^{2}[\textrm{SU(3)}]$ can cause complicated shape evolutions, which requres more investigations in future. If triaxiality dominates in the heavy nuclei, this principle would be even more important.}

In previous discussions, spectra and low-lying $B(E2)$ values of various $\gamma$-soft rotational modes in different nuclear structure theories are  very similar, so distinguishing various $\gamma$-softness in different theories is becoming extremely important. New $\gamma$-softness in the normal states of Cd nuclei can be only described by the SU3-IBM. However the spherical nucleus puzzle is still full of debates \cite{Garrett19,Garrett20,Nomura18,Leviatan18}. \textbf{A detailed study on $^{108-120}$ Cd with other SU(3) higher-order interactions have been recently performed  \cite{wang24}.} A complete description of $^{110,112}$Cd including configuration mixing, like Ref. \cite{Ramos11,Gavrielov19,Ramos19,Ramos20,Gavrielov22}, is in progress. $B(E2)$ values between higher levels are useful, such as the $0_{3}^{+}$, $2_{4}^{+}$ and $2_{5}^{+}$ states. Quadrupole moments of the low-lying states may provide great value in distinguishing among the various models.

Though a lot of theoretical work still needs to be done, we expect that many observations in $\gamma$-soft-like nuclei might be described by the SU3-IBM in a unified way. \textrm{E(5)}-like $\gamma$-softness in $^{82}$Kr is found in the new model \cite{zhou23}. Each interaction in the SU3-IBM has a clear geometric meaning, and different from common considerations in microscopic theory, such as SD-pair shell model \cite{Luo98,Zhao00}. In the microscopic theory, various deformations resulting from the proton-neutron interactions can give rise to different deformation shapes, including the $\gamma$-soft rotational mode \cite{Luo,He19}. In the SU3-IBM, this is not so, and specific interaction corresponds to certain shape, and the $\gamma$-softness is an emergent phenomenon, which can not be expected before numerical calculations. How to understand this point is an interesting problem. From the existing results, the SU3-IBM is closer to the geometric collective model \cite{Bohr75}. It seems important to investigate further the emerging $\gamma$-softness in the geometric model \cite{Fortunato05,Fortunato16}.

Since the seminal works by Elliott \cite{Elliott1,Elliott2,Harvey}, the \textrm{SU(3)} symmetry has played a key role for the description of the rotational spectra from the perspective of a microscopic shell model \cite{Kota20,Draayer1,Draayer2,Draayer3,Draayer4,Draayer5,Rowe1,Rowe2,Draayer6,Draayer7,Zuker1,Zuker2,Bonatsos171,Bonatsos23}. It was also found that the \textrm{SU(3)} symmetry plays a key role in the description of shell-like quartetting of nucleons, and the \textrm{SU(3)} third-order Casimir operator $\hat{C}_{3}[\textrm{SU(3)}]$ is needed to describe the experimental spectra in order to distinguish between the prolate and oblate shapes \cite{Cseh1,Cseh2,Cseh3,Cseh4}. In the previous studies, the \textrm{SU(3)} symmetry has been related to the prolate shape. In our new findings (\cite{Wang22,Wang20,Zhang22,wang23,zhou23,wangtao} and this paper), the \textrm{SU(3)} symmetry actually dominates all the quadrupole deformations.
In particular, in this model, one can obtain large quadrupole moments, even with spectra that resemble a gamma-soft situation.
We expect that this new perspective can be further used in the \textbf{pseudo-}\textrm{SU(3)} shell model \cite{Draayer4,Draayer5}.

 Along the dashed red line through the $A$ and $B$ point, shape phase transition from the prolate shape to the oblate shape has been studied \cite{wang23}, and this is an asymmetric evolution, which is different from the symmetric one in $\hat{H}_{1}$ \cite{Werner01,Jolie03}. The study of the properties of $^{196}$Pt in this paper further supports this perspective and a more detailed investigation on the prolate-oblate shape phase transition is also needed.

In addition, in Ref. \cite{zhou23}, the \textrm{E(5)}-like energy spectrum was also found in the new model, where there are also some new findings pointing to the $\gamma$-softness, which are further used to understand the $^{196}$Pt, which will be given in the second part (II) of the discussions \cite{zhou24}.

Rigid triaxial rotor in nuclear structure was first studied by Davydov and Filippov \cite{Davydov58}. In the SU(3) shell model, a SU(3) mapping of the rigid triaxial rotor can be obtained, which bridges between the geometrical model and the shell model \cite{Draayer87,Draayer881,Draayer882}. This SU(3) mapping was also discussed in the IBM \cite{Isacker00,zhang14}. Twenty years ago Wood \textit{et al.} investigated the triaxial rotor model by relaxing the use of irrotational moments of inertia \cite{Wood04}. Then this model was used to describe the $E2$ matrix elements available for $^{186,188,190,192}$Os \cite{Wood08}, which were usually treated as $\gamma$-soft nuclei. Importantly, the nearly zero value of the B(E2) transition from the $2_{2}^{+}$ state to the $0_{1}^{+}$ state in $^{196}$Pt can be explained by this model. These can help us further improve the fitting effect of $^{196}$Pt.

\textbf{SU3-IBM makes the SU(3) symmetry become even more important. In previous studies, the SU(3) symmetry is mainly related with the prolate shape and its rotational spectra \cite{Kota20}. In the SU3-IBM, the SU(3) symmetry has relationships with all the quadrupole deformations. Although this view is very simple, it was not realized before. In the SU(3) mapping of the rigid triaxial rotor, the SU(3) higher-order interactions are discussed \cite{Draayer87,Draayer881,Draayer882}, but they are not used to discuss the $\gamma$-softness in realistic nuclei. For the proxy-SU(3) symmetry, the energy shells for medium and heavy nuclei are reconsidered as the Elliott's SU(3) symmetry shells \cite{Bonatsos171,Bonatsos23}, thus the SU(3) symmetry also dominates the higher shells and the many-nucleons states can be described with the SU(3) symmetry. This is a breakthrough for understanding the single particle features. The proxy-SU(3) symmetry was also used to discuss the asymmetric prolate-oblate shape phase transition.  The reason of this phenomenon can be explained by the short-range nature of the nucleon-nucleon interaction, which leads to the dominance of the highest weight SU(3) irrep. We guess that, if the SU(3) higher-order interactions are considered here, new triaxial shape may be obtained. These results are somewhat consistent with our conclusions. The proxy-SU(3) symmetry seems to provide a microscopic foundation for the SU3-IBM. The proxy-SU(3)shell models with SU(3) higher-order interactions will be discussed in furture, and the mapping of the SU3-IBM from the proxy-SU(3) shell model should be also investigated.}

\textbf{The degree of $\gamma$-rigidity or $\gamma$-softness is determined by the fluctuation of the $\gamma$ degree of freedom, so it's hard to have a clear line between them. In our new study, it is found that $\gamma$-soft spectra can result from the $\gamma$-rigid ones \cite{zhou24}. In the geometric model, the Z(5) critical point symmetry was proposed to describe the critical nucleus from the prolate shape to the oblate shape \cite{Bonatsos04}, in which it is somewhat $\gamma$-rigid with $\gamma=30^{\circ}$. In addition, the equivalence between $\gamma$-soft and rigid triaxiality with $\gamma=30^{\circ}$ has been studied at the O(6) limit of IBM using projection techniques \cite{Otsuka87,Cohen88}. These questions require further study after our second paper \cite{zhou24}.}

In the SU(3) mapping, $[\hat{L} \times \hat{Q} \times \hat{L}]^{(0)}$, $[(\hat{L} \times \hat{Q})^{(1)} \times (\hat{L} \times \hat{Q})^{(1)}]^{(0)}$ and $\hat{L}^{2}$ are used to describe the rotational spectrum of the rigid triaxial rotor, but they can not affect the positions of the $0^{+}$ states. In this paper, it is shown that the $0^{+}$ states can be well explained by introducing the SU(3) third-order $\hat{C}_{3}[\textrm{SU(3)}]$ and the forth-order interactions $\hat{C}_{2}^{2}[\textrm{SU(3)}]$. Thus introducing the $[\hat{L} \times \hat{Q} \times \hat{L}]^{(0)}$ and $[(\hat{L} \times \hat{Q})^{(1)} \times (\hat{L} \times \hat{Q})^{(1)}]^{(0)}$ will further improve the fitting effect of the B(E2) values in Table I. A SU(3) mapping of the works done by Wood \textit{et al.} \cite{Wood04,Wood08,Wood10} will be first needed. These will be given in the third and \textbf{fourth} parts (III, IV) of the discussions.

\section{Conclusions}

Recently, the interacting boson model with \textrm{SU(3)} higher-order interactions (SU3-IBM) was proposed \cite{Wang22,Wang20,Zhang22} to resolve the spherical nucleus puzzle \cite{Garrett10,Garrett12,Batchelder12,Heyde16,Garrett18,Garrett19,Garrett20}, the $B(E2)$ anomaly \cite{168Os,166W,172Pt,170Os} and the prolate-oblate shape asymmetric evolution \cite{wang23,Werner01,Jolie03}. Although this model can be obtained by only considering the \textrm{U(5)} limit and the \textrm{SU(3)} limit, the $\gamma$-soft-like rotational mode can emerge. These results may extend our view of the interacting boson model (IBM). As said by Heyde and Wood: ``sphericity is a special case of deformation." and ``the reference frame must be fundamentally one of a deformed many-body system." \cite{Heyde16}. The emerging $\gamma$-softness may play a key role in the shift of perspective.

Following the previous studies \cite{Wang22,wang23}, the emerging $\gamma$-soft-like rotational mode can be explored to explain the properties of $^{196}$Pt. In our studies, a special point, which is near the middle point of the degenerate line connected the \textrm{U(5)} limit and the \textrm{SU(3)} degenerate point, is explored. The purpose of this paper is only to explore the relationship between the emerging $\gamma$-softness and the $\gamma$-soft properties in realistic nuclei. Further detailed fitting will be done in future when other \textrm{SU(3)} higher-order interactions are introduced, which is related to the destructive interference effect in the triaxial rotor model \cite{Wood04,Wood08,Wood10}.

Further investigation of the $\gamma$-rigid triaxiality is also important in the SU3-IBM. This is a delicate topic \cite{Nomura12}. The phase diagram of the SU3-IBM will be given in future, and it can offer meaningful guidance to the rigid triaxiality. 6-$d$ interaction may be also valuable \cite{Isacker81,Isacker84}. Distinguishing between the different $\gamma$-softness and discussing the differences between the $\gamma$-softness and the rigid triaxiality are topics that require further exploration.

Finally, a direct discussion and fitting of the new model on the oblate nuclei, such as $^{196-204}$Hg, is no doubt extremely important for understanding the SU3-IBM and for establishing the relationship between the new model and the \textrm{SU(3)} shell model. \textbf{And a discussion on the triaxiality of $^{166}$Er with SU3-IBM will be also considered in future.}

\section{ACKNOWLEDGMENT}

This research is supported by the Educational Department of Jilin Province, China (JJKH20210526KJ). C.-x.Z. gratefully acknowledges support from the Project Supported by
Scientific Research Fund of Hunan University of Art and Science (23ZZ04).

\end{document}